\PassOptionsToPackage{table}{xcolor}
\documentclass{article} 
\usepackage{iclr2025_conference,times}

\usepackage[T1]{fontenc}

\usepackage{amsmath,amsfonts,bm}

\def\eqref#1{equation~\ref{#1}}

\def\1{\bm{1}}

\DeclareMathAlphabet{\mathsfit}{\encodingdefault}{\sfdefault}{m}{sl}
\SetMathAlphabet{\mathsfit}{bold}{\encodingdefault}{\sfdefault}{bx}{n}

\usepackage{url}

\usepackage{amsmath,amssymb,amsfonts,amsthm,mathtools,bm}
\usepackage{enumitem}
\usepackage{multirow}
\usepackage{bm}
\usepackage{array}
\usepackage{color}
\usepackage[table]{xcolor} 
\definecolor{lightgray}{rgb}{0.88, 0.88, 0.88}
\definecolor{aliceblue}{rgb}{0.94, 0.97, 1.0}
\definecolor{blue(pigment)}{rgb}{0.2, 0.2, 0.6}
\newcolumntype{g}{>{\columncolor{lightgray}}}

\usepackage{wrapfig}
\usepackage{multicol}
\usepackage{caption}
\usepackage{diagbox}
\usepackage{pifont}
\usepackage{pdfpages}
\usepackage{arydshln}
\usepackage{booktabs}
\usepackage{makecell}
\usepackage{graphicx}
\usepackage{adjustbox}

\newcommand{\eg}{\textit{e}.\textit{g}.}
\newcommand{\ie}{\textit{i}.\textit{e}.}

\setcitestyle{authoryear,open={(},close={)}} 

\usepackage[colorlinks=true, linkcolor=blue(pigment), citecolor=blue(pigment)]{hyperref}
\usepackage[capitalise,noabbrev,nameinlink]{cleveref}
\crefformat{figure}{Figure~#2{\color{blue(pigment)}#1}#3}
\crefformat{table}{Table~#2{\color{blue(pigment)}#1}#3}
\crefformat{equation}{Equation~#2{\color{blue(pigment)}#1}#3}
\crefformat{section}{Section~#2{\color{blue(pigment)}#1}#3}
\crefformat{appendix}{Appendix~#2{\color{blue(pigment)}#1}#3}

\title{\system{}: Enhancing Ground-state Molecular Conformation Prediction via Force-Based Graph Rewiring}

\author{%
  Taewon Kim\textsuperscript{1}$^*$, 
  Hyunjin Seo \textsuperscript{1}$^*$, 
  Sungsoo Ahn \textsuperscript{2},
  Eunho Yang\textsuperscript{1,3}\\
  Korea Academic Insititue of Science and Technology (KAIST)\textsuperscript{1}, \\
  Pohang University of Science and Technology (POSTECH)\textsuperscript{2}, \\
  AITRICS\textsuperscript{3}\\
  \texttt{\{maxkim139,bella72,eunhoy\}@kaist.ac.kr,} \\
  \texttt{sungsoo.ahn@postech.ac.kr}
  }

\newcommand{\forceadj}{$\mathcal{A}^{\text{force}}$}
\newcommand{\attradj}{$\bm{A}^{\text{attr}}$}
\newcommand{\repadj}{$\bm{A}^{\text{rep}}$}

\begin{document}

\maketitle

\begin{abstract}
Predicting the ground-state 3D molecular conformations from 2D molecular graphs is critical in computational chemistry due to its profound impact on molecular properties. Deep learning (DL) approaches have recently emerged as promising alternatives to computationally-heavy classical methods such as density functional theory (DFT). However, we discover that existing DL methods inadequately model inter-atomic forces, particularly for non-bonded atomic pairs, due to their naive usage of bonds and pairwise distances. Consequently, significant prediction errors occur for atoms with low degree (\ie, low coordination numbers) whose conformations are primarily influenced by non-bonded interactions. To address this, we propose \system{}, a novel framework that rewires molecular graphs by adding edges based on the Lennard-Jones potential to capture non-bonded interactions for low-degree atoms. Experimental results demonstrate that \system{} significantly outperforms state-of-the-art methods across various molecular sizes, achieving up to a 20\% reduction in prediction error.
\end{abstract}

\section{Introduction}\label{1_introduction}
\def\thefootnote{*}\footnotetext{Equal contribution.}
The ground-state conformation of a molecule represents the lowest energy state on the potential energy surface, where the inter-atomic forces are balanced at equilibrium. This 3D conformation of the molecule plays a crucial role in determining the molecule’s physical, chemical, and biological properties. As a result, it is utilized in various applications such as molecular property prediction~\citep{satorras2021n, schutt2021equivariant, liu2022spherical, tholke2022torchmd, zhou2023unimol, zaidi2023pretraining, ni2024sliced}, drug discovery~\citep{luo2021predicting, ganea2021geomol, jing2022torsional, xu2022geodiff, zhou2023deep, tang2024survey}, and protein-ligand interactions~\citep{pei2024fabind, wang2024structure}.

Recently, deep learning approaches~\citep{gine, xu2021molecule3d, gatv2, graphgps, gtmgc} have emerged as promising alternatives to reduce the computational costs of ab initio calculations such as density functional theory \citep[DFT;][]{kohn1965self, parr1979local}. These approaches focus on predicting 3D molecular conformations by leveraging graph neural networks (GNNs) with 2D molecular graphs as input. Central to most of these models is the assumption that a 2D molecular graph, where atoms are represented as nodes and covalent bonds as edges, can effectively capture atomic interactions. Building on this assumption, they employ iterative message-passing updates to predict molecular conformations, which are parallel to the iterative force-based updates used for conformer optimization. Specifically, GNN-based methods rely on covalent bonds~\citep{gine, gatv2, graphgps} and pairwise distances~\citep{gtmgc} during the message-passing phase to represent inter-atomic forces, encouraging bonded and proximate atomic pairs to be close in the representational space.

In this work, we challenge the assumption of existing approaches that bonds and inter-atomic distances are sufficient to capture the complex behaviors of atomic interactions. 
In reality, forces are influenced by factors \textit{beyond} bonds and monotonic distance properties, especially for non-bonded atomic pairs that are typically dominated by van der Waals potentials \citep{vanderwaals, vanderwaals_nonbonddominance}. Consequently, significant prediction errors are observed for existing GNNs on atoms with low degree, \ie, low coordination numbers, whose conformations can be more sensitive to non-bonded interactions, as illustrated in \cref{fig:figure_analysis} in \cref{3_analysis}. Therefore, a new model to accurately account for the atomic interactions is warranted.


In response to this challenge, we present \system{}, a novel framework that rewires molecular graphs by adding edges between non-bonded atomic pairs exhibiting high inter-atomic forces, often modeled with the Lennard-Jones (LJ) potential~\citep{ljpotential}. To this end, we propose a novel encoder-decoder graph transformer architecture, where (1) the encoder generates an initial conformation prediction used for force-based rewiring and (2) the decoder predicts the final conformation from the rewired graph. 

To be specific, utilizing the encoder's prediction of inter-atomic distances from the initial conformation, \system{} calculates the absolute forces acting between all non-bonded atomic pairs through the derivative of the LJ potential. Subsequently, \system{} augments the graph by adding edges to non-bonded pairs with the largest computed forces. Our edge augmentation is performed in a degree-compensating manner, ensuring that atoms with fewer connections receive additional edges to enhance the modeling of its non-bonded interactions. Furthermore, to differentiate the nature of these interactions, we distinguish the augmented edges into distinct adjacency matrices each specifically modeling repulsive and attractive forces. Incorporating these force-based adjacency matrices, the decoder refines the initial conformation prediction by generating residual adjustments, leading to a more precise molecular geometry. 

The versatility of our proposed framework is demonstrated through benchmarks on both small-scale datasets, \ie, QM9~\citep{ramakrishnan2014quantum} and Molecule3D~\citep{xu2021molecule3d}, and a large-scale GEOM-DRUGS~\citep{axelrod2022geom} dataset. We also show that our idea of force-based rewiring brings universal improvements to GNNs even outside the proposed architecture, i.e., GINE~\citep{gine}, GATv2~\citep{gatv2}, and GraphGPS~\citep{graphgps}.

Our contributions are summarized as follows:
\begin{itemize}[topsep=0pt,itemsep=1mm, parsep=0pt, leftmargin=5mm]
\item We reveal that current approaches, which solely rely on bonds and predicted pairwise distances, are insufficient for accurately modeling inter-atomic forces, especially resulting in significant errors for nodes with low-degree atoms.
\item We introduce \system{}, a novel graph rewiring framework that adds edges between non-bonded atomic pairs with high inter-atomic forces guided by the Lennard-Jones potential. The number of augmented edges for each atom is determined in a degree-compensating fashion, improving the modeling of non-bonded interactions for low-degree atoms.
\item Extensive evaluation on diverse molecular sizes demonstrates the effectiveness of \system{} in enhancing ground-state molecular conformation prediction. Notably, our framework achieves improvements of up to 20\% on the QM9 dataset.
\end{itemize}

\section{Preliminaries}\label{2_preliminaries}

\paragraph{Problem definition.}\label{2.1}
We focus on predicting the 3D ground-state molecular conformation from its corresponding 2D molecular graph $\mathcal G=(\mathcal V, \mathcal E)$, where $\mathcal{V}$ denotes the set of $N = |\mathcal{V}|$ atoms (nodes) and $\mathcal{E}$ represents the set of $M = |\mathcal{E}|$ bonds (undirected edges) between atomic pairs. Nodes are characterized by a feature matrix $\bm X=[\bm x_1, \bm x_2, ..., \bm x_N]^{\mathsf T}\in\mathbb R^{N\times d}$, where each feature encodes atomic properties such as atom types and chirality. Edges are described by a binary adjacency matrix $\bm A\in\mathbb R^{N\times N}$, where $\bm A[i,j]=1$ if a bond exists between atoms $i$ and $j$, and $\bm A[i,j] = 0$ otherwise. Additionally, an edge feature matrix $\bm E=[\bm e_1, \bm e_2, ..., \bm e_M]^{\mathsf T}\in\mathbb R^{M\times f}$ may be utilized to represent bond-specific attributes, including bond types. The ground-state molecular conformation is represented as $\bm C=[\bm C_1, \bm C_2, ..., \bm C_N]^{\mathsf T}\in\mathbb R^{N\times 3}$, where each $\bm{C}_i \in \mathbb{R}^3$ corresponds to the 3D coordinate of atom $i$. The atomic pairwise distance is denoted as $\bm D\in\mathbb R^{N\times N}$, where $\bm D_{ij}=\lVert \bm C_i - \bm C_j\rVert_2$

\paragraph{Multi-head self-attention.}\label{2.2}
At the core of a transformer~\citep{vaswani2017attention} is a self-attention mechanism, which allows each instance in the input data to attend to every other instances, thereby enabling the model to capture the instance-wise relationships. Formally, given input feature matrix $\bm X$, the self-attention computes query ($\bm{Q}$), key ($\bm{K}$), and value ($\bm{V}$) matrices through linear transformations: 
\begin{equation} 
\bm{Q} = \bm{X} \bm{W}^Q, \quad \bm{K} = \bm{X} \bm{W}^K, \quad \bm{V} = \bm{X} \bm{W}^V, \end{equation} 
where $\bm{W}^Q$, $\bm{W}^K$, $\bm{W}^V \in \mathbb{R}^{d \times d_k}$ are learnable weight matrices. The attention scores at are then computed as follows: 
\begin{equation} 
 \text{Attention}(\bm{Q}, \bm{K}, \bm{V}) = \sigma_{\text{sm}}\left(\frac{\bm{Q} \bm{K}^\mathsf{T}}{\sqrt{d_k}}\right) \bm{V}, \end{equation} 
where $\sigma_{\text{sm}}$ is a softmax function and the scaling factor $\sqrt{d_k}$ stabilizes the gradients during training.

To enhance the model's ability to capture diverse patterns, multi-head attention employs multiple attention heads in parallel. Each head independently performs self-attention, and their outputs are concatenated and linearly transformed as:
\begin{equation} 
\text{MultiHead}(\bm{Q}, \bm{K}, \bm{V}) = \text{Concat}(\bm O_1, \dots, \bm O_H) \bm{W}^O, 
\end{equation}
where each $\bm O_h = \text{Attention}(\bm{Q}_h, \bm{K}_h, \bm{V}_h)$ and $\bm{W}^O \in \mathbb{R}^{Hd_k \times d_{\text{model}}}$ is a learnable weight matrix. 


\paragraph{Inter-atomic interaction modeling in prior works.}\label{2.3}
Traditional message-passing-based architectures~\citep{gine, gatv2, graphgps} leverage covalent bonds to model interactions between connected atomic pairs. For a given node $i$ at the $l$-th layer, the node representation $\bm h_i$ is updated via message aggregation from a set of its bonded neighboring nodes $\mathcal N_i$, as formulated below:
\begin{equation}
\bm h_{i}^{(l+1)}=\psi\bigg(\bm h_i^{(l)}, \;\phi\big(\{\bm h_j^{(l)},\bm g_{ij}^{(l)}\,|\,j\in\mathcal N_i\}\big)\bigg),
\end{equation}
where $\phi$ is an aggregation function that combines the representations of neighboring nodes, and $\psi$ is an update function that integrates this aggregated information with the node’s current hidden state. $\bm g_{ij}^{(l)}$ denotes the hidden representation of the edge feature corresponding to connected node pairs.

Recently, a new encoder-decoder based graph transformer~\citep{gtmgc} was proposed to capture inter-atomic forces by utilizing both bonds and atomic pairwise distances. In this framework, the pairwise distance matrix $\widehat{\bm D}\in\mathbb R^{N\times N}$ is computed from the initial conformation predicted by the encoder. For each $h$-th head in the decoder, the adjacency matrix $\bm A$ representing bond existence and a row-subtracted Euclidean distance matrix $\bm D^{\text{row-sub}}$, defined as $\bm D^{\text{row-sub}}_{ij}=\max_{n\in[1,N]}\widehat{\bm D}_{in}-\widehat{\bm D}_{ij}$, are integrated as residuals in the multi-head attention score $\bm{\widehat S}_{h}$, formulated as follows:
\begin{equation}\label{eqn_gtmgc}
    \begin{split}
        \bm{\widehat S}_{h} &= \bm S_h +  \bm S_{h}\odot (\beta_{h}^{A}\times \bm A) +  \bm S_{h}\odot (\beta_{h}^{D}\times \bm D^{\text{row-sub}}),\\
\bm S_{h}&=\bm Q_{h}\bm K_{h}^{\mathsf T}=(\bm Z\bm W_h^Q)(\bm Z\bm W_h^K)^\mathsf T,
    \end{split}
\end{equation}
where $\bm S_{h}\in\mathbb R^{N\times N}$ is the original self-attention score computed by the outer product between key and query matrices $\bm Q_h,\bm K_h\in\mathbb R^{N\times d_k}$, and $\beta_h^A,\beta_h^D\in\mathbb R$ are learnable parameters that determine the influence of bonds and distance factors, respectively.

\section{Limitation of inter-atomic force modeling in prior studies}\label{3_analysis}
In this section, we highlight the shortfalls of current approaches in using covalent bonds \citep{gine, gatv2, graphgps} or pairwise distances \citep{gtmgc} to model inter-atomic forces, followed by experimental results supporting our claim.

\begin{figure}[t!]
    \centering
    \includegraphics[width=\linewidth]{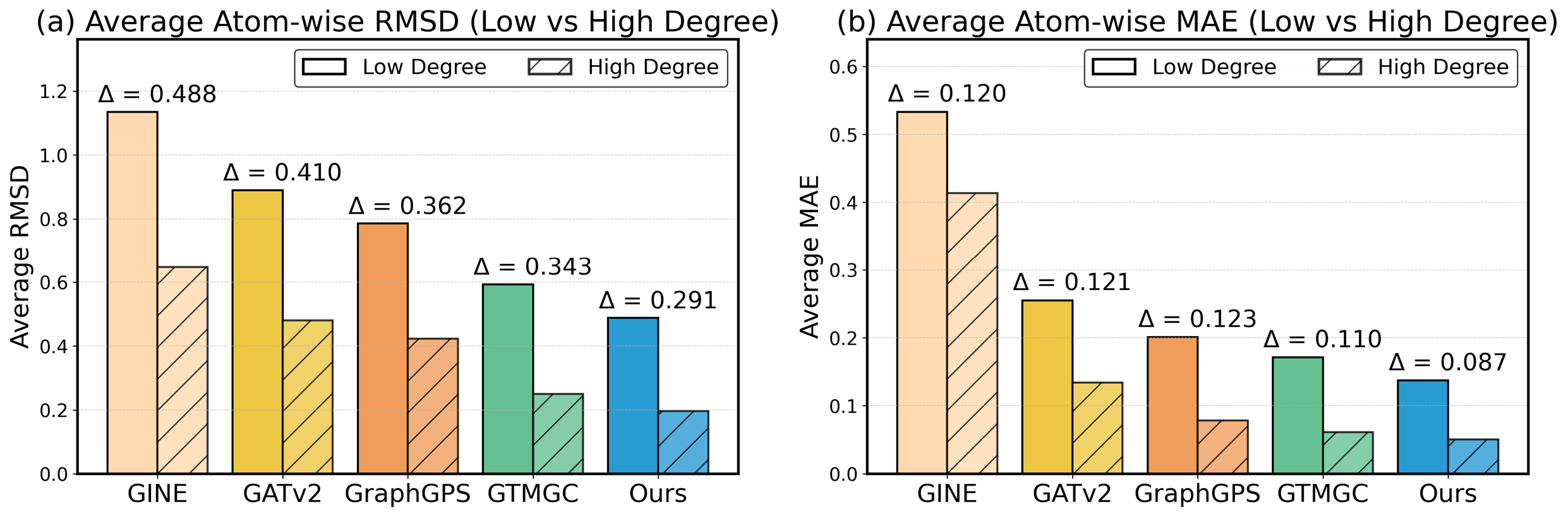}
    \vskip -5pt
    \caption{
   Atom-wise error analysis with respect to the relative atom degree on varing architectures in the QM9 dataset. (a) represents the average atom-wise RMSD while (b) shows the average atom-wise MAE for each bin of low-degree and high-degree atoms. $\Delta$ denotes the gap between the errors of the low-degree and high-degree groups.
    }
    \label{fig:figure_analysis}
    \vskip -10pt
\end{figure}

\paragraph{Limitations of prior works for non-bonded interactions.}
In ground-state molecular conformation prediction, accurately modeling inter-atomic forces is essential, as the net force on each atom defined by the gradient of molecular energy $E$, approaches zero at equilibrium~\citep{netforcezerodblp}. This equilibrium condition primarily determines the spatial arrangement of atoms. Meanwhile, a classic model of the energy $E$ divides the term into bonded and non-bonded interactions~\citep{moleculardynamics, luo2021predicting}, where bonded interactions can be further divided into bond stretching, angle bending, and torsion as shown in \cref{eqn_potential}.
\begin{equation}
\label{eqn_potential}
\begin{aligned}
E = E_{\text{bonded}} + E_{\text{non-bonded}} = {E}_{\text{bond}} + E_{\text{angle}} + E_{\text{torsion}} + E_{\text{non-bonded}}
\end{aligned}
\end{equation}
We find that previous works inadequately account for non-bonded interactions. As shown in \cref{2_preliminaries}, these methods incorporate the bond adjacency matrix $\bm{A}^{\text{bond}}$ to capture the bonded interactions $E_{\text{bonded}}$~\citep{gine, gatv2, graphgps}, or the distance-based proximity matrix $\bm{D}^{\text{row-sub}}$ (row-subtracted Euclidean distance matrix) to approximate non-bonded interactions $E_{\text{non-bonded}}$~\citep{gtmgc}. However, even the distance-based proximity matrix $\bm{D}^{\text{row-sub}}$ of GTMGC, which explicitly aims to model non-bonded interactions, oversimplifies non-bonded forces by assuming a monotonic relationship between attention weights and $\bm{D}^{\text{row-sub}}$. In reality, non-bonded interactions are governed by complex, non-linear functions of distance with atom type-specific coefficients~\citep{moleculardynamics}.

Consequently, this abstraction leads to performance degradation in atoms dominated by non-bonded potential $E_{\text{non-bonded}}$, and furthermore, $E$. We believe that atoms with fewer covalent bonds (\ie, low-degree atoms) correspond to such cases, due to being associated with fewer bonded potentials and hence being more sensitive to inaccuracies in the modeling of non-bonded interactions. We empirically verify this assertion in the subsequent paragraph.

\paragraph{Experimental verification.} 
To validate our assertion, we conduct an analysis on how prediction errors vary with respect to their node degree.
Specifically, we calculated an atom-wise Root Mean Square Deviation (RMSD) and Mean Absolute Error (MAE) of conformation predictions on the QM9 dataset. The calculation of atom-wise RMSD and MAE for each atom $i$ are as follows: 
\[
\text{RMSD}(i)=\lVert\widehat{\bm C}_i-\bm C_i\rVert_2,
\quad
\text{MAE}(i)=\frac{1}{N-1}\sum_{j\in \mathcal V\backslash \{i\}} |\widehat{\bm D}_{ij} - \bm D_{ij}|_1
\]
Subsequently, we categorize nodes into low-degree ($\text{deg}_i^{\text{rel}}\in(0, 0.3]$) and high-degree ($\text{deg}_i^{\text{rel}}\in[0.7, 1]$) groups based on their relative degree $\text{deg}_i^{\text{rel}}=\text{deg}_i^{(m)}/\max_{n\in[1,N]}\text{deg}_n^{(m)}$, then visualize the average errors for each group in \cref{fig:figure_analysis}. We evaluate upon GINE, GATv2 and GraphGPS which utilize covalent bond adjacency $\bm{A}^{\text{bond}}$ and GTMGC which additionally incorporates inter-atomic distance $\bm{D}^{\text{row-sub}}$.
As illustrated, the low-degree atoms exhibit significantly higher errors compared to high-degree atoms, with deviations up to 0.488 for $\text{RMSD}(i)$ and 0.123 for $\text{MAE}(i)$.   This demonstrates that previous works fail to sufficiently model inter-atomic forces between non-bonded atomic pairs, hindering the effective modeling of $E_{\text{non-bonded}}$.

\begin{figure}[t!]
    \centering
    \includegraphics[width=\linewidth]{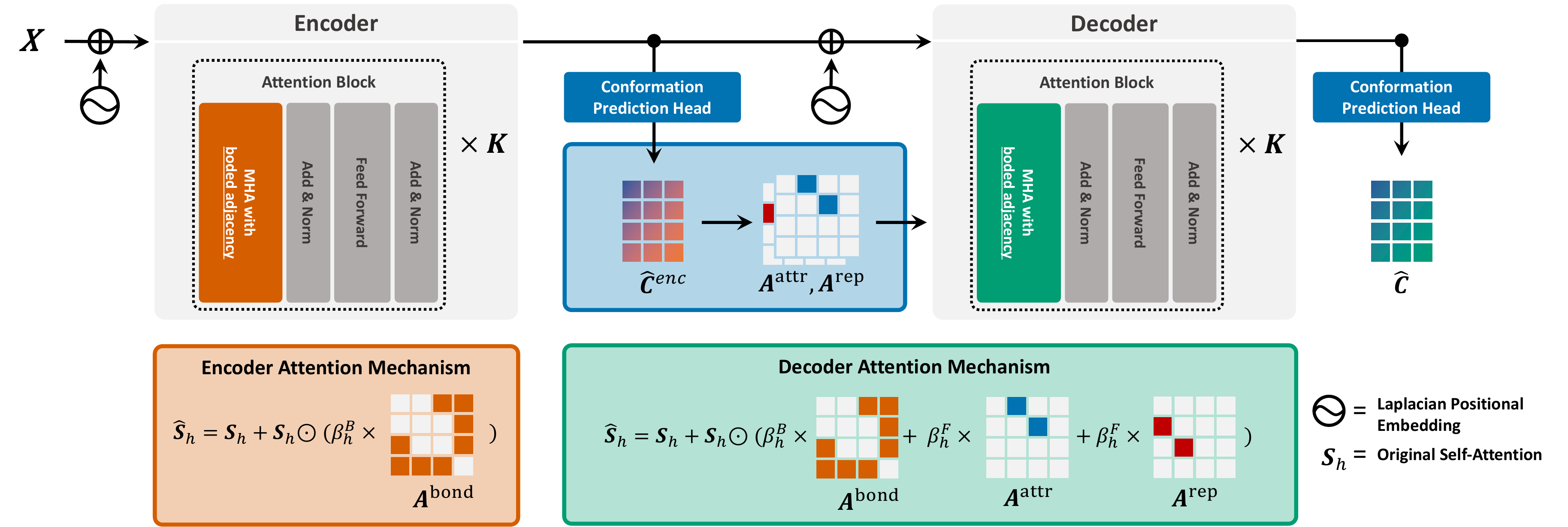}
    \caption{
    Overview of the \system{} framework.  
    }
    \label{Fig: figure_main}
    \vskip -10pt
\end{figure}

\section{REBIND: Enhancing ground-state molecular conformation via force-based graph rewiring}\label{4_proposed_method}
Given the limitation of previous studies, we introduce \system{}, a novel graph rewiring framework that selectively adds edges in a force-aware manner, prioritizing atoms with low-degree. In \cref{4.1}, we provide an overview of the \system{} architecture. \cref{4.2} details the force-aware graph rewiring component, and \cref{4.3} describes the integration of the augmented edges into the multi-head self-attention. Finally, the learning objective of \system{} and comparison with existing studies are detailed in \cref{4.4}.

\subsection{Overview}\label{4.1}
The overall architecture of \system{} is illustrated in \cref{Fig: figure_main}. Our framework receives a 2D molecular graph as input, characterized by its bonded adjacency matrix $\bm{A}^{\text{bond}}$ and node feature or embeddings augmented with Laplacian positional encoding $\bm L$. The framework outputs a predicted molecular conformation, denoted as $\bm{\widehat{C}}$. We employ a standard encoder-decoder architecture with multi-head self-attention, following the approaches presented in \citep{vaswani2017attention, cai2020graph, gtmgc}. In this setup, the encoder processes the input graph to generate hidden representations $\bm{H}^{\text{enc}}$. A task-specific prediction head then produces an initial conformation prediction $\widehat{\bm C}^{\text{enc}}$ from $\bm{H}^{\text{enc}}$. From this intermediate prediction, we derive 1) an inter-atomic distance matrix $\widehat{\bm{D}}^{\text{enc}}$ and 2) force-aware adjacency matrices $\mathcal{A}^{\text{force}} = \{\bm{A}^{\text{attr}}, \bm{A}^{\text{rep}}\}$, where \attradj~and \repadj~denote adjacency matrices identifying attractive and repulsive atomic pairs. The hidden representation $\bm H^{\text{enc}}$, along with $\bm A^{\text{bond}}$ and \forceadj, are subsequently fed into the decoder to generate the refined, residual molecular representation.


\subsection{Force-aware graph rewiring}\label{4.2}
Here, we outline the construction of force-aware adjacency matrix \forceadj, which connects non-bonded atomic pairs exerting significant forces to one another, mitigating the shortfalls of prior works identified in \cref{3_analysis}.

\paragraph{Force modeling with Lennard-Jones potential.} To capture the interactions between non-bonded atomic pairs, we leverage the Lennard-Jones (LJ) potential~\citep{ljpotential}, a well-established model for non-bonded interactions such as van der Waals forces. The LJ potential is defined as:
\begin{equation}
    V(r) = 4 \varepsilon \left[ \left( \frac{\sigma}{r} \right)^{12} - \left( \frac{\sigma}{r} \right)^{6} \right],
\end{equation}
where $r$ is the distance between two atoms, $\varepsilon$ is the depth of the potential well (representing the strength of the interaction), and $\sigma$ is the finite distance at which the inter-atomic potential is zero. The first term $\left( \frac{\sigma}{r} \right)^{12}$ accounts for the repulsion when atoms are too close, while the second term $\left( \frac{\sigma}{r} \right)^{6}$ accounts for the attraction at moderate distances. By using the derivative of this potential, we compute the inter-atomic forces acting upon non-bonded atom pairs, which is critical for modeling the equilibrium conformation of the molecule. This force model provides a more realistic representation of how spatial positions are influenced, especially for low-degree atoms, where non-covalent interactions dominate.

The values of $\sigma$ and $\varepsilon$ are assigned based on the specific pair of atom types being modeled (\eg, carbon, hydrogen, etc.). In line with conventional force-field modeling approaches used in computational chemistry, we adopt the predefined parameter values from the Universal Force Field (UFF)~\citep{rappe1992uff}. For interactions between different atom types, the parameters are determined using the Lorentz-Berthelot mixing rules. Specifically, the interaction parameters $\sigma_{ij}$ and $\varepsilon_{ij}$ for atom pair $i$ and $j$ are defined as follows:
\begin{equation}
    \sigma_{ij} = \frac{\sigma_i + \sigma_j}{2},\; \varepsilon_{ij} = \sqrt{\varepsilon_i \cdot \varepsilon_j}
\end{equation}
Thus, the full inter-atomic force between atom $i$ and $j$ can be written as:
\begin{equation}\label{eqn_force}
F(r) = -\frac{dV(r)}{dr} =  24 \varepsilon_{ij} \left[ 2 \left( \frac{\sigma_{ij}}{r} \right)^{12} - \left( \frac{\sigma_{ij}}{r} \right)^{6} \right] \frac{1}{r}
\end{equation}
Here, the sign of $F(r)$ denotes the directionality of the force, where positive values correspond to repulsion, and negative values indicate attraction.

\begin{figure}[t!]
    \centering
    \includegraphics[width=.9\linewidth]{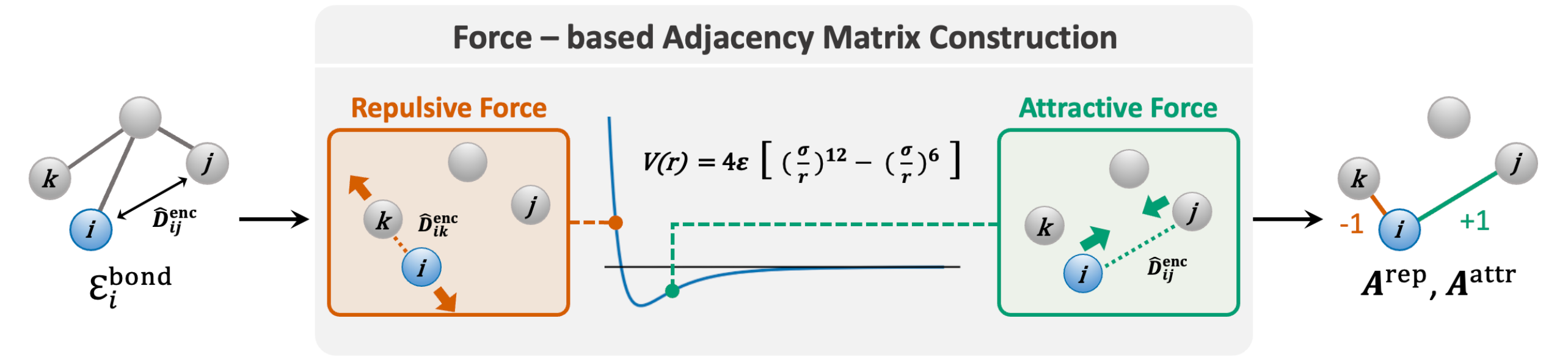}
    \vskip -5pt
    \caption{
    An illustration of edge augmentation in \system{} using the LJ potential. Non-bonded atomic pairs with the largest force magnitudes are added as edges to the molecular graph in a degree-compensating manner. The augmented edges are treated according to the nature of the forces, distinguishing between attractive and repulsive interactions.
    }
    \label{Fig: figure_edge_augmentation}
    \vskip -10pt
\end{figure}

\paragraph{Construction of force adjacency matrix.}
Given the derived force function from the LJ potential, we construct a set of force adjacency matricesdj \forceadj~ which introduces additional edges to atoms that significantly impact its spatial conformation, while prioritizing those of low-degree. Utilizing the predicted distance matrix obtained from the encoder $\widehat{\bm{D}}^{\text{enc}}$, where 
$\widehat{\bm{D}}^{\text{enc}}_{ij} = \|\bm{C}_i - \bm{C}_j\|_2 = \sqrt{(\bm{C}_i - \bm{C}_j)^{\mathsf T}(\bm{C}_i - \bm{C}_j)}$, we compute the inter-atomic force  $\bm{F}^{\text{LJ}}$ for non-bonded atomic pairs using the \cref{eqn_force}. 
We then augment the graph by connecting each node $i$ to the top $\text{K}_i=\max_{n \in [1, N]} \text{deg}_n^{m} - \text{deg}i^{m}$ non-bonded atoms that exert the largest forces on $i$ as follows:
\begin{equation} 
\mathcal{E}^{\text{force}}_i = \left\{ j \ \middle| \ j \in \text{Top}_{\text{K}_i}\left( \{j \ |\;|\bm{F}^{\text{LJ}}[i,j]|\}\right), \ \bm{e}_{ij} \notin \mathcal{E}^{\text{bond}} \right\}
\end{equation}
Here, for each atom $i$, we select non-bonded neighbors $j$ that exert the largest forces on $i$, adding up to the maximum possible degree $\max_{n\in[1,N]}\text{deg}_n^{m} - \text{deg}_j^{m}$ without thresholding or introducing any hyperparameters. To differentiate the nature of these forces, we decompose $\mathcal{E}^{\text{force}}_{i}$ into separate adjacency matrices $\bm{A}^{\text{attr}}$ and $\bm{A}^{\text{rep}}$ which capture attraction and repulsion, respectively. With $\mathbb{1}$ as the indicator function, we formalize $\bm{A}^{\text{attr}}$ and $\bm{A}^{\text{rep}}$ as follows:
\[
    \bm{A}^{\text{attr}}_{ij} = \mathbb{1}\left[ (i,j) \in \mathcal{E}^{\text{force}}_i \wedge \bm{F}^{\text{LJ}}[i,j] < 0 \right],   \bm{A}^{\text{rep}}_{ij} = -\mathbb{1}\left[ (i,j) \in \mathcal{E}^{\text{force}}_i \wedge \bm{F}^{\text{LJ}}[i,j] \geq 0 \right]
\]
Note that the edge weights of $\bm{A}^{attr}$ are positive ones (\ie~$+1$) while weights of $\bm{A}^{rep}$ are negative ones (\ie~$-1$), ensuring distinctiveness between the directionality of forces.
The complete process is outlined in \cref{Fig: figure_edge_augmentation}.

\subsection{Integration with multi-head self-attention}\label{4.3}
Leveraging the augmented edges, \system{} enhance the multi-head self-attention by incorporating both bond-based and force-based adjacency matrices. Specifically, the augmented force-based adjacency matrices $\mathcal{A}^{\text{force}} = \{\bm{A}^{\text{attr}}, \bm{A}^{\text{rep}}\}$ and original bond-based adjacency matrix $\bm{A}^{\text{bond}}$ are incorporated as residual components in the attention scores for each head $h$. This integration is formalized as follows:
\begin{equation}
    \bm{\widehat S}_{h} = \bm S_h +  \bm S_{h}\odot (\beta_{h}^{B}\times \bm A^{\text{bond}}) +  \bm S_{h}\odot (\beta_{h}^{Attr}\times \bm A^{\text{attr}}) +  \bm S_{h}\odot (\beta_{h}^{Rep}\times \bm A^{\text{rep}}),
\end{equation}
where $\bm S_{h}=\bm Q_{h}\bm K_{h}^{\mathsf T}=(\bm Z\bm W_h^Q)(\bm Z\bm W_h^K)^\mathsf T$ represents the global self-attention score. Here, $\bm Z=\bm H^{\text{enc}}+\bm L\in\mathbb R^{N\times d}$ is the input to the decoder, combining the encoder's hidden representations $\bm{H}^{\text{enc}}$ with Laplacian positional encoding $\bm{L}$. $\beta_h^B,\beta_h^{Attr},\beta_h^{Rep}$ are learnable parameters that modulate the influence of 1) bonds, 2) attraction and 3)repulsion adjacency matrices, respectively.
Subsequently, the output of each attention head, $\bm{O}_h$, is computed by applying a softmax normalization to the scaled attention scores and then multiplying by the linear projection of the decoder input $\bm{V}_h \in \mathbb{R}^{N \times d_k}$:
\begin{equation}
    \begin{split}
        \bm O_h &= \sigma_{\text{sm}}\bigg(\frac{\bm{\widehat S}_{h}}{\sqrt{d_k}}\bigg)\bm V_h,\quad
        \bm V_h = \bm Z\bm W_h^V,
    \end{split}
\end{equation}
where $\sigma_{\text{sm}}$ is a softmax function applied to the attention score scaled by $\sqrt{d_k}$. The final conformation $\widehat{\bm C}$ is obtained by refining the initial encoder representation $\bm H^{\text{enc}}$ with the residual decoder representation $\bm H^{\text{dec}}$, which is achieved via channel-wise attention:
\begin{equation}
\begin{split}
\widehat{\bm{C}} &= \text{FFN}(\bm Y)\in\mathbb{R}^{N \times 3}, \quad
\bm Y = \sum_{i=1}^2\bm\alpha_{:,:,i}\odot\bm H_{:,:,i}\in\mathbb{R}^{N \times d_c}, \\
\bm \alpha &= \sigma_{\text{sm}}\left(\frac{\bm H^{\text{enc}}\bm W_y\;||\;\bm H^{\text{dec}}\bm W_y}{\sqrt{d_c}}\right) \in \mathbb{R}^{N \times d_c \times 2}, \quad
\bm H = \bm{H}^{\text{enc}} \; || \; \bm{H}^{\text{dec}} \in\mathbb{R}^{N \times d_{\text{model}}\times 2},
\end{split}
\end{equation}
where $\text{FFN}(\cdot)$ is a task-specific head layer, $\bm W_y\in\mathbb R^{_{\text{model}}\times d_c}$ is a linear transformation applied to the hidden representations, and $||$ denotes concatenation along the last dimension. The channel-wise attention weights $\bm{\alpha}$ determine the contribution of the encoder and decoder representations in producing the final conformation.

Throughout this attention mechanism, \system{} effectively models inter-atomic forces by learning from both bonded and non-bonded atomic pairs. By incorporating the structural supervision from $\bm{A}^{\text{bond}}$ and the force-aware supervision from \forceadj~ into the global attention scores $\bm{S}$, \system{} facilitates comprehensive force modeling, enhancing the accuracy of conformation predictions.

\subsection{Objective function}\label{4.4}
To ensure that the predicted molecular conformation remains invariant to rotation and translation, we adopt a loss function based on the difference between the predicted and ground-truth pairwise Euclidean atomic distances, denoted as $\widehat{\bm{D}}$ and $\bm{D}$, respectively, which is also utilized in prior works \citep{xu2021molecule3d, gtmgc}. Additionally, to achieve precise force modeling within the decoder, we apply the same loss objective to the initial conformation predictions $\widehat{\bm{D}}^{\text{enc}}$ and $\bm{D}$ from the encoder.
The overall loss objective $\mathcal L$ is formulated as follows:
\begin{equation}
    \mathcal L = \frac{1}{N^2}\sum_i^N\sum_j^N|\widehat{\bm D}_{ij}^{\text{enc}}-\bm D_{ij}| + \frac{1}{N^2}\sum_i^N\sum_j^N|\widehat{\bm D}_{ij}-\bm D_{ij}|
\end{equation}

\paragraph{Comparison with existing graph transformers.} Several existing works integrate inter-atomic relationships as residuals within multi-head self-attention mechanisms. For instance, the Geometric Transformer \citep{choukroun2021geometric} utilizes the pairwise Euclidean distances $\bm{D}$, computed from ground-truth atomic coordinates, to perform molecular property prediction. Similarly, MAT \citep{maziarka2020molecule} incorporates both the pairwise distances $\bm{D}$ and the bond adjacency matrix $\bm{A}^{\text{bond}}$ for the same task. In contrast, GTMGC \citep{gtmgc} employs predicted inter-atomic distances $\bm{D}^{\text{row-sub}}$ to forecast ground-state molecular conformations.

While our work also leverages pairwise atomic relationships as residuals in the computation of attention scores, \system{} distinguishes itself by enabling fine-grained force modeling. This is achieved through the joint utilization of both bonded interactions and non-bonded interactions characterized by the largest forces. Consequently, our framework attains more precise ground-state molecular geometries, which is substantiated in the subsequent section.

\renewcommand{\arraystretch}{1.2}
  \begin{table*}[t!]
  \aboverulesep=0ex 
   \belowrulesep=0ex 
    \caption{Conformer prediction performance and percentage reduction (\%) of \system{} compared to the best baseline performance on the QM9 and Molecule3D datasets.}
    \vskip -3pt
    \centering
    \resizebox{\textwidth}{!}{%
    \huge
    \begin{tabular}{ c | c | c c c c | c c c c }
    \toprule[2.2pt]
     \rowcolor{lightgray}\multicolumn{2}{ c |}{\fontsize{24}{24}\selectfont\textbf{Splits}} & \multicolumn{4}{| c |}{\fontsize{24}{24}\selectfont Validation} &
     \multicolumn{4}{| c }{\fontsize{24}{24}\selectfont Test} \\
     \midrule[2.2pt]
     \textbf{\fontsize{24}{24}\selectfont Datasets} & \textbf{\fontsize{24}{24}\selectfont Methods} 
     & \fontsize{24}{24}\selectfont D-MAE $\downarrow$ & \fontsize{24}{24}\selectfont D-RMSE $\downarrow$ 
     & \fontsize{24}{24}\selectfont C-RMSD $\downarrow$ & \fontsize{24}{24}\selectfont E-RMSD $\downarrow$ 
     
     & \fontsize{24}{24}\selectfont D-MAE $\downarrow$ & \fontsize{24}{24}\selectfont D-RMSE $\downarrow$ 
     & \fontsize{24}{24}\selectfont C-RMSD $\downarrow$ & \fontsize{24}{24}\selectfont E-RMSD $\downarrow$ \\
     \hline
     \multirow{9}{*}{\textbf{QM9}} 
     & RDKit-DG & 
     \fontsize{24}{24}\selectfont 0.328 & 
     \fontsize{24}{24}\selectfont 0.570 &
     \fontsize{24}{24}\selectfont 0.502 &
     \fontsize{24}{24}\selectfont 1.044 &
     
     \fontsize{24}{24}\selectfont 0.330 &
     \fontsize{24}{24}\selectfont 0.573 &
     \fontsize{24}{24}\selectfont 0.504 &
     \fontsize{24}{24}\selectfont 1.266 \\
     
     & RDKit-ETKDG & 
     \fontsize{24}{24}\selectfont 0.324 & 
     \fontsize{24}{24}\selectfont 0.574 &
     \fontsize{24}{24}\selectfont 0.458 &
     \fontsize{24}{24}\selectfont 1.048 &
     
     \fontsize{24}{24}\selectfont 0.325 &
     \fontsize{24}{24}\selectfont 0.574 &
     \fontsize{24}{24}\selectfont 0.460 &
     \fontsize{24}{24}\selectfont 1.120 \\
     
     & GINE & 
     \fontsize{24}{24}\selectfont 0.605 & 
     \fontsize{24}{24}\selectfont 0.950 &
     \fontsize{24}{24}\selectfont 0.865 &
     \fontsize{24}{24}\selectfont 1.703 &
     
     \fontsize{24}{24}\selectfont 0.606 &
     \fontsize{24}{24}\selectfont 0.946 &
     \fontsize{24}{24}\selectfont 0.867 &
     \fontsize{24}{24}\selectfont 1.696 \\
     
     & GATv2 & 
     \fontsize{24}{24}\selectfont 0.382 &
     \fontsize{24}{24}\selectfont 0.695 &
     \fontsize{24}{24}\selectfont 0.711 &
     \fontsize{24}{24}\selectfont 1.371 &
     
     \fontsize{24}{24}\selectfont 0.382 &
     \fontsize{24}{24}\selectfont 0.690 &
     \fontsize{24}{24}\selectfont 0.712 &
     \fontsize{24}{24}\selectfont 1.358 \\
     
     & GraphGPS (RW) & 
     \fontsize{24}{24}\selectfont 0.328 &
     \fontsize{24}{24}\selectfont 0.629 &
     \fontsize{24}{24}\selectfont 0.628 &
     \fontsize{24}{24}\selectfont 1.196 &
     
     \fontsize{24}{24}\selectfont 0.327 &
     \fontsize{24}{24}\selectfont 0.624 &
     \fontsize{24}{24}\selectfont 0.628 &
     \fontsize{24}{24}\selectfont 1.193 \\
     
     & GraphGPS (LP) & 
     \fontsize{24}{24}\selectfont 0.283 &
     \fontsize{24}{24}\selectfont 0.500 &
     \fontsize{24}{24}\selectfont 0.544 &
     \fontsize{24}{24}\selectfont 1.049 &
     
     \fontsize{24}{24}\selectfont 0.283 &
     \fontsize{24}{24}\selectfont 0.499 &
     \fontsize{24}{24}\selectfont 0.546 &
     \fontsize{24}{24}\selectfont 1.064 \\
     
     & GTMGC & 
     \fontsize{24}{24}\selectfont 0.280 &
     \fontsize{24}{24}\selectfont 0.470 &
     \fontsize{24}{24}\selectfont 0.415 &
     \fontsize{24}{24}\selectfont 0.792 &
     
     \fontsize{24}{24}\selectfont 0.281 &
     \fontsize{24}{24}\selectfont 0.471 &
     \fontsize{24}{24}\selectfont 0.414 &
     \fontsize{24}{24}\selectfont 0.800 \\
     
     & \fontsize{24}{24}\selectfont \system{} & 
     \cellcolor{aliceblue}\textbf{\fontsize{24}{24}\selectfont 0.252} &
     \cellcolor{aliceblue}\textbf{\fontsize{24}{24}\selectfont 0.442} &
     \cellcolor{aliceblue}\textbf{\fontsize{24}{24}\selectfont 0.320} &
     \cellcolor{aliceblue}\textbf{\fontsize{24}{24}\selectfont 0.601} &
     
     \cellcolor{aliceblue}\textbf{\fontsize{24}{24}\selectfont 0.254} &
     \cellcolor{aliceblue}\textbf{\fontsize{24}{24}\selectfont 0.446} &
     \cellcolor{aliceblue}\textbf{\fontsize{24}{24}\selectfont 0.321} &
     \cellcolor{aliceblue}\textbf{\fontsize{24}{24}\selectfont 0.610} \\
     \cdashline{2-10}
     
     & {\fontsize{24}{24}\selectfont {Reduction $\uparrow$}} & 
     \fontsize{24}{24}\selectfont 10.00 &
     \fontsize{24}{24}\selectfont 5.96 &
     \fontsize{24}{24}\selectfont 22.89 &
     \fontsize{24}{24}\selectfont 24.12 &
     
     \fontsize{24}{24}\selectfont 9.61 &
     \fontsize{24}{24}\selectfont 5.31 &
     \fontsize{24}{24}\selectfont 22.46 &
     \fontsize{24}{24}\selectfont 23.75 \\
     \hline
     
     \multirow{10}{*}{\textbf{\makecell{Molecule3D \\(random)}}} 
     & RDKit-DG & 
     \fontsize{24}{24}\selectfont 0.581 & 
     \fontsize{24}{24}\selectfont 0.930 &
     \fontsize{24}{24}\selectfont 1.043 &
     \fontsize{24}{24}\selectfont 1.864 &
     
     \fontsize{24}{24}\selectfont 0.582 &
     \fontsize{24}{24}\selectfont 0.932 &
     \fontsize{24}{24}\selectfont 1.044 &
     \fontsize{24}{24}\selectfont 1.872 \\
     
     & RDKit-ETKDG & 
     \fontsize{24}{24}\selectfont 0.575 & 
     \fontsize{24}{24}\selectfont 0.942 &
     \fontsize{24}{24}\selectfont 0.981 &
     \fontsize{24}{24}\selectfont 1.700 &
     
     \fontsize{24}{24}\selectfont 0.576 &
     \fontsize{24}{24}\selectfont 0.943 &
     \fontsize{24}{24}\selectfont 0.983 &
     \fontsize{24}{24}\selectfont 1.710 \\
     
     & DeeperGCN-DAGNN & 
     \fontsize{24}{24}\selectfont 0.509 & 
     \fontsize{24}{24}\selectfont 0.849 &
     \fontsize{24}{24}\selectfont N/A &
     \fontsize{24}{24}\selectfont N/A &
     
     \fontsize{24}{24}\selectfont 0.571 &
     \fontsize{24}{24}\selectfont 0.961 &
     \fontsize{24}{24}\selectfont N/A &
     \fontsize{24}{24}\selectfont N/A \\
     
     & GINE & 
     \fontsize{24}{24}\selectfont 0.591 &
     \fontsize{24}{24}\selectfont 1.016 &
     \fontsize{24}{24}\selectfont 1.103 &
     \fontsize{24}{24}\selectfont 2.227 &
     
     \fontsize{24}{24}\selectfont 0.592 &
     \fontsize{24}{24}\selectfont 1.019 &
     \fontsize{24}{24}\selectfont 1.104 &
     \fontsize{24}{24}\selectfont 2.230 \\
     
     & GATv2 & 
     \fontsize{24}{24}\selectfont 0.564 &
     \fontsize{24}{24}\selectfont 0.985 &
     \fontsize{24}{24}\selectfont 1.072 &
     \fontsize{24}{24}\selectfont 2.163 &
     
     \fontsize{24}{24}\selectfont 0.565 &
     \fontsize{24}{24}\selectfont 0.989 &
     \fontsize{24}{24}\selectfont 1.073 &
     \fontsize{24}{24}\selectfont 2.168 \\
     
     & GraphGPS (RW) & 
     \fontsize{24}{24}\selectfont 0.512 &
     \fontsize{24}{24}\selectfont 0.900 &
     \fontsize{24}{24}\selectfont 1.006 &
     \fontsize{24}{24}\selectfont 2.089 &
     
     \fontsize{24}{24}\selectfont 0.513 &
     \fontsize{24}{24}\selectfont 0.903 &
     \fontsize{24}{24}\selectfont 1.008 &
     \fontsize{24}{24}\selectfont 2.094 \\
     
     & GraphGPS (LP) & 
     \fontsize{24}{24}\selectfont 0.440 &
     \fontsize{24}{24}\selectfont 0.730 &
     \fontsize{24}{24}\selectfont 0.854 &
     \fontsize{24}{24}\selectfont 1.687 &
     
     \fontsize{24}{24}\selectfont 0.441 &
     \fontsize{24}{24}\selectfont 0.732 &
     \fontsize{24}{24}\selectfont 0.854 &
     \fontsize{24}{24}\selectfont 1.689 \\
     
     & GTMGC & 
     \fontsize{24}{24}\selectfont 0.429 &
     \fontsize{24}{24}\selectfont 0.713 &
     \fontsize{24}{24}\selectfont 0.708 &
     \fontsize{24}{24}\selectfont 1.347 &
     
     \fontsize{24}{24}\selectfont 0.430 &
     \fontsize{24}{24}\selectfont 0.715 &
     \fontsize{24}{24}\selectfont 0.709 &
     \fontsize{24}{24}\selectfont 1.350 \\
     
     & \fontsize{24}{24}\system{} & 
     \cellcolor{aliceblue}\textbf{\fontsize{24}{24}\selectfont 0.418} &
     \cellcolor{aliceblue}\textbf{\fontsize{24}{24}\selectfont 0.706} &
     \cellcolor{aliceblue}\textbf{\fontsize{24}{24}\selectfont 0.698} &
     \cellcolor{aliceblue}\textbf{\fontsize{24}{24}\selectfont 1.314} &
     
     \cellcolor{aliceblue}\textbf{\fontsize{24}{24}\selectfont 0.419} &
     \cellcolor{aliceblue}\textbf{\fontsize{24}{24}\selectfont 0.708} &
     \cellcolor{aliceblue}\textbf{\fontsize{24}{24}\selectfont 0.699} &
     \cellcolor{aliceblue}\textbf{\fontsize{24}{24}\selectfont 1.317} \\
     \cdashline{2-10}

     & {\fontsize{24}{24}\selectfont {Reduction $\uparrow$}} & 
     \fontsize{24}{24}\selectfont 2.56 &
     \fontsize{24}{24}\selectfont 0.98 &
     \fontsize{24}{24}\selectfont 1.41 &
     \fontsize{24}{24}\selectfont 2.45 &
     
     \fontsize{24}{24}\selectfont 2.56 &
     \fontsize{24}{24}\selectfont 0.98 &
     \fontsize{24}{24}\selectfont 1.41 &
     \fontsize{24}{24}\selectfont 2.44 \\
     \hline
     
     \multirow{10}{*}{\textbf{\makecell{Molecule3D \\(scaffold)}}} 
     & RDKit-DG & 
     \fontsize{24}{24}\selectfont 0.542 & 
     \fontsize{24}{24}\selectfont 0.872 &
     \fontsize{24}{24}\selectfont 0.993 &
     \fontsize{24}{24}\selectfont 1.751 &
     
     \fontsize{24}{24}\selectfont 0.524 &
     \fontsize{24}{24}\selectfont 0.857 &
     \fontsize{24}{24}\selectfont 0.970 &
     \fontsize{24}{24}\selectfont 1.780 \\
     
     & RDKit-ETKDG & 
     \fontsize{24}{24}\selectfont 0.531 & 
     \fontsize{24}{24}\selectfont 0.874 &
     \fontsize{24}{24}\selectfont 0.916 &
     \fontsize{24}{24}\selectfont 1.565 &
     
     \fontsize{24}{24}\selectfont 0.511 &
     \fontsize{24}{24}\selectfont 0.858 &
     \fontsize{24}{24}\selectfont 0.892 &
     \fontsize{24}{24}\selectfont 1.595 \\
     
     & DeeperGCN-DAGNN & 
     \fontsize{24}{24}\selectfont 0.617 & 
     \fontsize{24}{24}\selectfont 0.930 &
     \fontsize{24}{24}\selectfont N/A &
     \fontsize{24}{24}\selectfont N/A &
     
     \fontsize{24}{24}\selectfont 0.763 &
     \fontsize{24}{24}\selectfont 1.176 &
     \fontsize{24}{24}\selectfont N/A &
     \fontsize{24}{24}\selectfont N/A \\
     
     & GINE & 
     \fontsize{24}{24}\selectfont 0.889 &
     \fontsize{24}{24}\selectfont 1.517 &
     \fontsize{24}{24}\selectfont 1.398 &
     \fontsize{24}{24}\selectfont 3.007 &
     
     \fontsize{24}{24}\selectfont 1.388 &
     \fontsize{24}{24}\selectfont 2.200 &
     \fontsize{24}{24}\selectfont 1.928 &
     \fontsize{24}{24}\selectfont 4.142 \\
     
     & GATv2 & 
     \fontsize{24}{24}\selectfont 0.791 &
     \fontsize{24}{24}\selectfont 1.402 &
     \fontsize{24}{24}\selectfont 1.264 &
     \fontsize{24}{24}\selectfont 2.675 &
     
     \fontsize{24}{24}\selectfont 1.248 &
     \fontsize{24}{24}\selectfont 2.082 &
     \fontsize{24}{24}\selectfont 1.768 &
     \fontsize{24}{24}\selectfont 3.832 \\
     
     & GraphGPS (RW) & 
     \fontsize{24}{24}\selectfont 0.503 &
     \fontsize{24}{24}\selectfont 0.853 &
     \fontsize{24}{24}\selectfont 0.978 &
     \fontsize{24}{24}\selectfont 2.047 &
     
     \fontsize{24}{24}\selectfont 0.595 &
     \fontsize{24}{24}\selectfont 1.024 &
     \fontsize{24}{24}\selectfont 1.065 &
     \fontsize{24}{24}\selectfont 2.203 \\
     
     & GraphGPS (LP) & 
     \fontsize{24}{24}\selectfont 0.417 &
     \fontsize{24}{24}\selectfont 0.690 &
     \fontsize{24}{24}\selectfont 0.827 &
     \fontsize{24}{24}\selectfont 1.639 &
     
     \fontsize{24}{24}\selectfont 0.411 &
     \fontsize{24}{24}\selectfont 0.690 &
     \fontsize{24}{24}\selectfont 0.827 &
     \fontsize{24}{24}\selectfont 1.606 \\
     
     & GTMGC & 
     \fontsize{24}{24}\selectfont 0.406 &
     \fontsize{24}{24}\selectfont 0.670 &
     \fontsize{24}{24}\selectfont 0.682 &
     \fontsize{24}{24}\selectfont 1.318 &
     
     \fontsize{24}{24}\selectfont 0.397 &
     \fontsize{24}{24}\selectfont 0.671 &
     \fontsize{24}{24}\selectfont 0.692 &
     \fontsize{24}{24}\selectfont 1.275 \\
     
     & \fontsize{24}{24}\selectfont \system{} & 
     \cellcolor{aliceblue}\textbf{\fontsize{24}{24}\selectfont 0.391} &
     \cellcolor{aliceblue}\textbf{\fontsize{24}{24}\selectfont 0.661} &
     \cellcolor{aliceblue}\textbf{\fontsize{24}{24}\selectfont 0.640} &
     \cellcolor{aliceblue}\textbf{\fontsize{24}{24}\selectfont 1.174} &
     
     \cellcolor{aliceblue}\textbf{\fontsize{24}{24}\selectfont 0.386} &
     \cellcolor{aliceblue}\textbf{\fontsize{24}{24}\selectfont 0.663} &
     \cellcolor{aliceblue}\textbf{\fontsize{24}{24}\selectfont 0.667} &
     \cellcolor{aliceblue}\textbf{\fontsize{24}{24}\selectfont 1.182} \\
     \cdashline{2-10}

     & {\fontsize{24}{24}\selectfont {Reduction $\uparrow$}} & 
     \fontsize{24}{24}\selectfont 3.69 &
     \fontsize{24}{24}\selectfont 4.20 &
     \fontsize{24}{24}\selectfont 6.16 &
     \fontsize{24}{24}\selectfont 10.93 &
     
     \fontsize{24}{24}\selectfont 2.77 &
     \fontsize{24}{24}\selectfont 1.19 &
     \fontsize{24}{24}\selectfont 3.61 &
     \fontsize{24}{24}\selectfont 7.29 \\

    \bottomrule[2.2pt]
    \end{tabular}}
    \label{Tab: table_main}
    \vskip -10pt
  \end{table*}

\section{Experiments}
\label{5_experiments}

\subsection{Experimental setup}\label{5.1}
\paragraph{Datasets.} 
We evaluated \system{} on well-established benchmark datasets, including QM9~\citep{ramakrishnan2014quantum}, Molecule3D~\citep{xu2021molecule3d}, and GEOM-DRUGS~\citep{axelrod2022geom}. QM9 consists of small organic molecules and is widely utilized for quantum chemistry applications. Molecule3D is a large-scale dataset of molecular structures, for which we employed two distinct splitting strategies: random split and scaffold split. The scaffold split groups molecules based on their core substructures, enabling a more realistic evaluation. Additionally, GEOM-DRUGS comprises large-size molecules relevant to drug discovery, providing a challenging benchmark for assessing the scalability of our framework on complex molecular structures. 
Since the original dataset includes multiple stable conformations, we choose the most stable conformation with respect to the Boltzmann energy for each molecule. Detailed descriptions of each dataset are provided in \cref{C}.

\paragraph{Metrics.}
Following previous work~\citep{gtmgc}, we evaluate the performance of our model using Mean Absolute Error (MAE), Root Mean Squared Error (RMSE), and Root Mean Square Deviation (RMSD). Additionally, we introduce a new metric, Energy-weighted RMSD (E-RMSD), which accounts for the chemical feasibility of predicted conformations. By denoting $\bm G_i$ and $\bm G$ as the \textit{aligned} predicted and ground-state conformation via the Kabsch algorithm (\cite{kabsch}), E-RMSD is calculated as:
\begin{equation}
    \text{E-RMSD}(\widehat{\bm{G}}, \bm{G}) = \frac{p}{\widehat{p}}\sqrt{\sum_{i\in\mathcal V}{w}_i\lVert\widehat{\bm{G}}_i - \bm{G}_i\rVert_2}
\end{equation}
Here, $\frac{p}{\widehat{p}}$ denotes the Boltzmann factor denoted as  $\text{exp}\left(\frac{\hat{E} - E}{kT}\right)$, while atom-wise normalized force $w_i$ is defined as $w_i = \frac{F_i}{\sum_{j\in\mathcal V}{F_{j}}}$. $E$ and $\hat{E}$ denotes the total energy of ground-truth and predicted conformations, and $F_i$ denotes the force calculated via the force-field defined from the predicted conformation, both computed using the Merck Molecular Force Field~\citep{mmff}. The constants $k$ and $T$ are each the Boltzmann constant and thermodynamic temperature, respectively. This approach penalizes 1) errors of molecules that are energetically unstable via $\frac{p}{\widehat{p}}$, and 2) errors of atoms that are off-equilibrium with respect to net force via $w_i$. Detailed descriptions of other metrics are provided in \cref{C.2}

\paragraph{Baselines.}
Following \cite{gtmgc}, we compared our \system{} against traditional cheminformatic methods, DG and ETKDG algorithms, from RDKit~\citep{landrum2013rdkit} and five representative GNNs. The GNNs considered in our experiments can be broadly broadly categorized as (1) traditional GNNs such as GINE~\citep{gine}, GATv2~\citep{gatv2}, DeeperGCN-DAGNN adapted from \cite{xu2021molecule3d}, (2) graph transformers including GraphGPS~\citep{graphgps} and GTMGC~\citep{gtmgc}. For GraphGPS, we conducted experiments using both random walk (RW) and Laplacian (LP) positional encoding strategies. 
To ensure a fair comparison, we trained the MoleBERT~\citep{xia2023mole} tokenizer separately for each dataset, which differs from the original setting where the tokenizer trained on a Molecule3D random split is used universally across all benchmarks.
To further validate the efficacy of our framework, we included a comparison with the diffusion-based conformer generation model, Torsional Diffusion~\citep{jing2022torsional}.
Detailed experimental configurations are specified in \cref{C}.

\subsection{Main results}\label{5.2}
We present the performance results with of \system{} in \cref{Tab: table_main}, where the percentage reduction is calculated as the ratio of the performance improvement over the best baseline, expressed as a percentage.
As demonstrated, our method achieves consistent superiority over baseline methods across all datasets and evaluation splits. On the QM9 dataset, \system{} achieves notable improvements across all evaluation metrics, with gains of up to 24\% in both C-RMSD and E-RMSD. Additionally, our method exhibits robust generalization capabilities on the larger Molecule3D dataset. Notably, \system{} further excels in the scaffold split, where training and test sets contain structurally diverse molecules, achieving up to a 10\% reduction in E-RMSD. This highlights \system{}'s ability to generalize effectively to novel molecular scaffolds.
Furthermore, it is worth emphasizing that higher performance gains are observed in E-RMSD compared to C-RMSD. This suggests that \system{} generates molecular conformations that are not only geometrically accurate but also more \textit{realistic} in terms of energetic stability, resulting in more physically plausible molecular structures.

\renewcommand{\arraystretch}{1.65}
  \begin{wraptable}{r}{0.55\textwidth}
  \vskip -13pt
  \aboverulesep=0ex 
   \belowrulesep=0ex 
    \caption{Conformer prediction performance and percentage reduction (\%) of \system{} compared to the best baseline performance on the GEOM-DRUGS dataset.}
    \vskip -3pt
    \centering
    \resizebox{.56\textwidth}{!}{%
    \Huge
    \begin{tabular}{ c | c | c c c c }
    \toprule[.7pt]
     \rowcolor{lightgray}\multicolumn{2}{ c |}{\fontsize{35}{35}\selectfont\textbf{Splits}} & 
     \multicolumn{4}{| c }{\fontsize{35}{35}\selectfont Test} \\
     \midrule[.7pt]
     \textbf{\fontsize{35}{35}\selectfont Datasets} & \textbf{\fontsize{35}{35}\selectfont Methods} 
     
     & \fontsize{34}{34}\selectfont D-MAE $\downarrow$ & \fontsize{34}{34}\selectfont D-RMSE $\downarrow$ 
     & \fontsize{34}{34}\selectfont C-RMSD $\downarrow$ & \fontsize{34}{34}\selectfont E-RMSD $\downarrow$ \\
     \hline
     \multirow{9}{*}{\fontsize{35}{35}\selectfont \textbf{GEOM-DRUGS}} 
     & \fontsize{35}{35}\selectfont RDKit-DG & 
     \fontsize{36}{36}\selectfont 1.181 &
     \fontsize{36}{36}\selectfont 2.132 &
     \fontsize{36}{36}\selectfont 2.097 &
     \fontsize{36}{36}\selectfont 3.623 \\
     
     & \fontsize{35}{35}\selectfont RDKit-ETKDG & 
     \fontsize{36}{36}\selectfont 1.120 &
     \fontsize{36}{36}\selectfont 2.055 &
     \fontsize{36}{36}\selectfont 1.934 &
     \fontsize{36}{36}\selectfont 3.330 \\
     
     & \fontsize{35}{35}\selectfont GINE & 
     \fontsize{36}{36}\selectfont 1.125 &
     \fontsize{36}{36}\selectfont 1.777 &
     \fontsize{36}{36}\selectfont 2.033 &
     \fontsize{36}{36}\selectfont 3.925 \\
     
     & \fontsize{35}{35}\selectfont GATv2 & 
     \fontsize{36}{36}\selectfont 1.042 &
     \fontsize{36}{36}\selectfont 1.662 &
     \fontsize{36}{36}\selectfont 1.901 &
     \fontsize{36}{36}\selectfont 3.728 \\
     
     & \fontsize{35}{35}\selectfont GraphGPS (RW) & 
     \fontsize{36}{36}\selectfont 0.879 &
     \fontsize{36}{36}\selectfont 1.399 &
     \fontsize{36}{36}\selectfont 1.768 &
     \fontsize{36}{36}\selectfont 3.472 \\
     
     & \fontsize{35}{35}\selectfont GraphGPS (LP) & 
     \fontsize{36}{36}\selectfont 0.815 &
     \fontsize{36}{36}\selectfont 1.300 &
     \fontsize{36}{36}\selectfont 1.698 &
     \fontsize{36}{36}\selectfont 3.171 \\
     
     & \fontsize{35}{35}\selectfont GTMGC & 
     \fontsize{36}{36}\selectfont 0.823 &
     \fontsize{36}{36}\selectfont 1.319 &
     \fontsize{36}{36}\selectfont 1.458 &
     \fontsize{36}{36}\selectfont 2.830 \\

     & \fontsize{35}{35}\selectfont Torsional Diffusion & 
     \fontsize{36}{36}\selectfont 0.959 &
     \fontsize{36}{36}\selectfont 1.648 &
     \fontsize{36}{36}\selectfont 1.751 &
     \fontsize{36}{36}\selectfont 2.992 \\
     
     & \fontsize{35}{35}\selectfont \system{} & 
     \cellcolor{aliceblue}\textbf{\fontsize{36}{36}\selectfont 0.776} &
     \cellcolor{aliceblue}\textbf{\fontsize{36}{36}\selectfont 1.283} &
     \cellcolor{aliceblue}\textbf{\fontsize{36}{36}\selectfont 1.396} &
     \cellcolor{aliceblue}\textbf{\fontsize{36}{36}\selectfont 2.602} \\
     \cdashline{2-6}

     & {\fontsize{35}{35}\selectfont {Percentage Reduction $\uparrow$}} & 
     \rule[1.2ex]{0pt}{-.3ex}
     \fontsize{36}{36}\selectfont 4.79 &
     \fontsize{36}{36}\selectfont 1.31 &
     \fontsize{36}{36}\selectfont 4.25 &
     \fontsize{36}{36}\selectfont 8.06 \\

    \bottomrule[.7pt]
    \end{tabular}}
    \label{Tab: table_drugs}
    \vskip -12pt
  \end{wraptable}

\subsection{Scalability to large molecules}\label{5.3}
We further validate \system{} on the GEOM-DRUGS dataset with large molecules by including Torsional Diffusion~\citep{jing2022torsional}as an additional baseline, which adopts the same dataset as its benchmark. Since Torsional Diffusion was originally designed to generate multiple conformers, we adapt it to produce a single conformation per molecule for evaluation. 

As shown in \cref{Tab: table_drugs}, \system{} significantly outperforms the baselines, achieving improvements of up to 4\% in C-RMSD and 8\% in E-RMSD. Diffusion-based models like Torsional Diffusion are designed to generate multiple stable conformations and are not directly suited for predicting the single most stable conformation of a molecule, resulting in suboptimal performance. In contrast, \system{} demonstrates strong generalizability to drug-like molecules, surpassing diffusion-based models with lower computational costs and eliminating the need for multiple inference steps.

\subsection{Ablative study}\label{5.4}
\renewcommand{\arraystretch}{1}
\begin{wraptable}{r}{0.55\textwidth}
\vskip -13pt
\aboverulesep=0ex 
   \belowrulesep=0ex 
    \caption{Ablative study of \system{}.}
    \vskip -4pt
    \centering
    \resizebox{.55\textwidth}{!}{
    \huge
    \begin{tabular}{ >{\centering}p{2.5cm}  >{\centering}p{2.5cm}  >{\centering}p{2.5cm} >{\centering}p{2.5cm} | >{\centering}p{2.5cm} | >{\centering}p{2.5cm} }
    \toprule[.7pt]
     {\fontsize{16}{16}\selectfont $\bm L$} & 
     {\fontsize{16}{16}\selectfont $\bm A^{\text{bond}}$} &
     {\fontsize{16}{16}\selectfont $\bm A^{\text{near}}$} &
     {\fontsize{16}{16}\selectfont $\mathcal A^{\text{force}}$} & {\fontsize{15}{15}\selectfont C-RMSD $\downarrow$} &
     {\fontsize{15}{15}\selectfont E-RMSD $\downarrow$} \tabularnewline
     \midrule[.7pt]
     \fontsize{16}{16}\selectfont $\checkmark$
     &  
     &  
     &  
     &  \fontsize{16}{16}\selectfont 0.5294
     &  \fontsize{16}{16}\selectfont 1.0323 \tabularnewline

     \fontsize{16}{16}\selectfont \checkmark
     & \fontsize{16}{16}\selectfont \checkmark
     &  
     &  
     &  \fontsize{16}{16}\selectfont 0.3524
     &  \fontsize{16}{16}\selectfont 0.6920 \tabularnewline

     \fontsize{16}{16}\selectfont \checkmark
     & \fontsize{16}{16}\selectfont \checkmark
     & \fontsize{16}{16}\selectfont \checkmark
     &  
     &  \fontsize{16}{16}\selectfont 0.3522
     &  \fontsize{16}{16}\selectfont 0.6807 \tabularnewline

     \rowcolor{aliceblue}\fontsize{16}{16}\selectfont \checkmark
     & \fontsize{16}{16}\selectfont \checkmark
     &  
     & \fontsize{16}{16}\selectfont \checkmark
     &  \fontsize{16}{16}\selectfont \textbf{0.3209}
     &  \fontsize{16}{16}\selectfont \textbf{0.6103} \tabularnewline
     \bottomrule[.7pt]
    \end{tabular}}
    \label{Tab:table_ablation}
    \vskip -10pt
\end{wraptable}
\paragraph{Ablation on components.} We conduct an ablation study on the components of \system{} on the QM9 dataset, with the results presented in \cref{Tab:table_ablation}. 
Using a fully-connected transformer with Laplacian positional embedding $\bm{L}$ results in limited performance, highlighting that uniformly connecting edges without accounting for inter-atomic forces is suboptimal.
Integrating the bond-based adjacency matrix $\bm{A}^{\text{bond}}$ with $\bm{L}$ leads to substantial performance improvements.
Moreover, we evaluate the impact of incorporating force-aware edges (\ie, \forceadj) compared to edge augmentation based on proximity (\ie, $\bm{A}^{\text{near}}$), the strategy adopted in \cite{luo2021predicting}. The distance-based edges $\bm{A}^{\text{near}}$ are constructed using the intermediate distance matrix $\widehat{\bm{D}}^{\text{enc}}$, connecting atoms $i$ and $j$ when ${\widehat{\bm{D}}^{\text{enc}}}_{ij}<\delta$. The threshold $\delta$ is set as $10\,\text{\AA}$, following the protocol in~\cite{luo2021predicting}. 
Our results demonstrate that \forceadj, which accounts for significant inter-atomic forces  
influencing spatial conformation, consistently outperforms $\bm{A}^{\text{near}}$, showing up to 9\% improvements in C-RMSD and larger gains up to 10\% in E-RMSD. This demonstrates the effectiveness of \forceadj~ in aiding the model to generate conformations that are both closer to the ground state and energetically stable.

\paragraph{Application to GNN-based architectures.} 
The core concept of \system{} is adaptable to any encoder-decoder framework. To demonstrate its versatility, we integrate \system{} with three GNN-based architectures: GINE, GATv2, and GraphGPS (LP). We implement \system{} by dividing the $L$ layers of original GNNs into two halves, where the top half operates as the encoder and the bottom half operates as the decoder. 
To compute \forceadj~, we follow the same procedure as \system{}, wherein the encoder’s output is passed through a task-specific head layer to generate $\widehat{\bm{C}}^{\text{enc}}$ and $\widehat{\bm{D}}^{\text{enc}}$. 

The results are shown in \cref{Tab:table_gnn_ablation_test} of \cref{B}. Results demonstrate that incorporating \system{} significantly improves performance across all backbone architectures and metrics, with gains of at least 5\%. 
Notably, when applied to GraphGPS, we achieve the lowest prediction errors, with improvements of 5.11\% in C-RMSD and 8.74\% in E-RMSD. These results underscore the versatility of \system{} in boosting performance regardless of the underlying architecture. Moreover, \system{}’s integration requires minimal implementation effort, making it compatible with any existing GNN-based architecture through the proposed mechanism.




\subsection{Qualitative analysis}\label{5.5}
We present a qualitative comparison of \system{} against baseline architectures based on atom-wise RMSD on the QM9 dataset, shown as blue bars in \cref{fig:figure_analysis}. 
As illustrated, \system{} achieves significant error reductions for both groups.
Specifically, when compared against GTMGC, \system{} further reduces the atom-wise RMSD for lower-degree groups by 0.11, compared to a reduction of 0.05 for high-degree groups, thereby verifying the efficacy of our degree-compensating augmentation.
Moreover, our method greatly reduces the gap between the atom-wise errors of low-degree and high-degree groups, achieving 15.16\% and 20.91\% percentage reduction in the gap for atom-wise RMSD and MAE, respectively, compared to GTMGC. 
These results substantiates that \system{} comprehensively captures non-bonded interactions through force-aware edge augmentation.

\section{Related works}
\label{6_related_works}
There has been a growing interest in generating molecular conformations from 2D molecular graphs. Traditional cheminformatics methods, such as DG and ETKDG from RDKit~\citep{landrum2013rdkit}, are widely used for their efficiency, relying on chemical heuristics for fast generation. However, their performance is limited due to insufficient handling of non-bonded interactions and minimal energy optimization.
To address this limitation, deep learning models have emerged as powerful alternatives, producing multiple conformations for molecules using generative models~\citep{mansimov2019molecular, xu2021end, xu2021learning, luo2021predicting, shi2021learning, xu2022geodiff, jing2022torsional}. 
Recently, the focus has shifted to predicting ground-state molecular conformations, which is critical for practical applications requiring stability and feasibility. A benchmark and training pipeline were introduced in \cite{xu2021molecule3d} for this task, followed by the graph transformer~\citep{gtmgc}, achieving state-of-the-art results. Further details on related works are provided in \cref{A}.

\section{Conclusion}
\label{7_conclusion}
Given the limitations of previous studies in modeling inter-atomic forces, we introduced \system{}, an innovative framework that incorporates force-aware edge augmentation to accurately capture inter-atomic interactions. 
We anticipate that \system{} can be extended to other critical applications such as drug discovery or property prediction, particularly in scenarios where ground-truth coordinates are unavailable. As a future work, we plan to integrate our framework with these applications to enhance their predictive accuracy.


\bibliography{iclr2025_conference}
\bibliographystyle{iclr2025_conference}

\newpage
\appendix
\section*{Supplementary Materials}

\renewcommand{\arraystretch}{1.1}
  \begin{table}[h]
  \aboverulesep=0ex 
   \belowrulesep=0ex 
    \caption{Conformer prediction performance and percentage reduction (\%) of GNN-based backbones integrated with \system{} on the QM9 dataset.}
    \vskip -3pt
    \centering
    \resizebox{.7\textwidth}{!}{%
    \huge
    \begin{tabular}{ c | c | c c c c }
    \toprule[2pt]
     \rowcolor{lightgray}\multicolumn{2}{ c |}{\fontsize{23}{23}\selectfont\textbf{Method}} & 
     \multicolumn{4}{| c }{\fontsize{23}{23}\selectfont Test} \\
     \midrule[1.8pt]
     \textbf{\fontsize{23}{23}\selectfont Backbone} & \textbf{\fontsize{23}{23}\selectfont Methods} 
     & \fontsize{23}{23}\selectfont D-MAE $\downarrow$ & \fontsize{23}{23}\selectfont D-RMSE $\downarrow$ 
     & \fontsize{23}{23}\selectfont C-RMSD $\downarrow$ & \fontsize{23}{23}\selectfont E-RMSD $\downarrow$ \\
     \midrule[.7pt]
     \multirow{3}{*}{\textbf{GINE}} 
     & Vanilla 
     & \fontsize{23}{23}\selectfont 0.606 
     & \fontsize{23}{23}\selectfont 0.946 
     & \fontsize{23}{23}\selectfont 0.867 
     & \fontsize{23}{23}\selectfont 1.696 \\
     
     &+ \system{}
     & \cellcolor{aliceblue}\textbf{\fontsize{23}{23}\selectfont 0.505} 
     & \cellcolor{aliceblue}\textbf{\fontsize{23}{23}\selectfont 0.761} 
     & \cellcolor{aliceblue}\textbf{\fontsize{23}{23}\selectfont 0.801} 
     & \cellcolor{aliceblue}\textbf{\fontsize{23}{23}\selectfont 1.605} \\
     \cdashline{2-6}
     & Reduction $\uparrow$ 
     & \fontsize{23}{23}\selectfont 16.65 
     & \fontsize{23}{23}\selectfont 19.55 
     & \fontsize{23}{23}\selectfont 7.61 
     & \fontsize{23}{23}\selectfont 5.36 \\
     \midrule[.7pt]
     
     \multirow{3}{*}{\textbf{GATv2}} 
     & Vanilla 
     & \fontsize{23}{23}\selectfont 0.382 
     & \fontsize{23}{23}\selectfont 0.690 
     & \fontsize{23}{23}\selectfont 0.712 
     & \fontsize{23}{23}\selectfont 1.358 \\
     
     &+ \system{}
     & \cellcolor{aliceblue}\textbf{\fontsize{23}{23}\selectfont 0.319} 
     & \cellcolor{aliceblue}\textbf{\fontsize{23}{23}\selectfont 0.523} 
     & \cellcolor{aliceblue}\textbf{\fontsize{23}{23}\selectfont 0.611} 
     & \cellcolor{aliceblue}\textbf{\fontsize{23}{23}\selectfont 1.198} \\
    \cdashline{2-6}
     & Reduction $\uparrow$ 
     & \fontsize{23}{23}\selectfont 16.62 
     & \fontsize{23}{23}\selectfont 24.14 
     & \fontsize{23}{23}\selectfont 14.14 
     & \fontsize{23}{23}\selectfont 11.77 \\
     \midrule[.7pt]
     
     \multirow{3}{*}{\textbf{\makecell{GraphGPS \\(LP)}}} 
     & Vanilla 
     & \fontsize{23}{23}\selectfont 0.283 
     & \fontsize{23}{23}\selectfont 0.499 
     & \fontsize{23}{23}\selectfont 0.546 
     & \fontsize{23}{23}\selectfont 1.064 \\
     
     &+ \system{}
     & \cellcolor{aliceblue}\textbf{\fontsize{23}{23}\selectfont 0.211} 
     & \cellcolor{aliceblue}\textbf{\fontsize{23}{23}\selectfont 0.361} 
     & \cellcolor{aliceblue}\textbf{\fontsize{23}{23}\selectfont 0.518} 
     & \cellcolor{aliceblue}\textbf{\fontsize{23}{23}\selectfont 0.971} \\
     \cdashline{2-6}
     & Reduction $\uparrow$ 
     & \fontsize{23}{23}\selectfont 25.55 
     & \fontsize{23}{23}\selectfont 27.58 
     & \fontsize{23}{23}\selectfont 5.11 
     & \fontsize{23}{23}\selectfont 8.74 \\
     \bottomrule[1.8pt]
    \end{tabular}}
    \label{Tab:table_gnn_ablation_test}
    \vskip -10pt
  \end{table}

\section{Related Works}\label{A}
\paragraph{Molecular Conformer Generation.}
In recent years, there has been significant progress in the field of molecular conformation generation from 2D molecular graphs. Traditional cheminformatics approaches, such as Distance Geometry (DG) and its extension, ETKDG~\citep{landrum2013rdkit}, have been widely employed due to their computational efficiency. These methods rely on chemical heuristics and geometric rules to generate conformations efficiently, which makes them attractive for large-scale applications. However, their reliance on approximations and minimal energy optimization limits their ability to accurately capture non-bonded interactions, which play a crucial role in determining the stability and realism of molecular conformations.

To overcome the limitations of traditional approaches, deep learning models have emerged as promising choices. Generative models, in particular, have been extensively explored for molecular conformation generation, offering more accurate and realistic predictions. 
Early works~\citep{mansimov2019molecular, xu2021end} leverage variational autoencoders (VAEs) to encode molecular information into latent spaces, where each molecule's structure is represented as a probabilistic distribution. From this latent space, multiple plausible 3D conformations are sampled by decoding the latent variables. Similarly, flow-based models~\citep{xu2021learning} generate molecular conformations by transforming latent variables sampled from a Gaussian distribution into atomic distance matrices, capturing long-range dependencies between atoms. The 3D coordinates are then derived from the generated distances, followed by refinement using energy-based models.
Score-based models have also gained attention for molecular conformation generation. These models learn a score function representing the gradient of the log probability density of atomic coordinates~\citep{luo2021predicting, shi2021learning}. By perturbing the molecular data with Gaussian noise at multiple levels, these models iteratively denoise the data using the learned score function to guide the generation of valid conformations.
Building on the success of diffusion models~\citep{ho2020denoising} in the vision domain, their application to molecular conformation generation has become another emerging area of interest~\citep{xu2022geodiff, jing2022torsional, zhang2023sdegen, fan2024ec}. They have shown remarkable success in generating diverse molecular geometries by learning the desired geometric distribution from a noise distribution through a reverse diffusion process.

\paragraph{Ground-state molecular conformer prediction.}
While the one-to-many task of generating multiple conformations has been thoroughly studied, recent research has shifted toward the prediction of a molecule's ground-state conformation. This focus reflects the practical importance of identifying the most stable and energetically favorable molecular structure, which is essential for real-world applications such as drug design and material discovery. 
Ground-state conformation prediction requires models to identify the most stable configuration on the molecule's potential energy surface, which is challenging for one-to-many methods to address.

To facilitate progress in this area, the Molecule3D benchmark and training pipeline are introduced in \cite{xu2021molecule3d}. They provide a standardized dataset and evaluation protocols for predicting a molecule's ground-state geometry. Building on this, the graph transformer architecture (GTMGC)~\citep{gtmgc} was recently proposed, tailored to predict ground-state molecular conformations. GTMGC incorporates bonds and pairwise distances within the self-attention mechanism to capture both local and global molecular interactions, achieving state-of-the-art performance on the benchmarks with ground-state conformations.

\renewcommand{\arraystretch}{3.6}
\setlength\arrayrulewidth{3.2pt}
\begin{table*}[ht!]
\aboverulesep=0ex 
   \belowrulesep=0ex 
\centering
\caption{Examples of visualizations and RMSD predictions from RDKit-ETKDG, GTMGC, and \system{} on the QM9 and GEOM-DRUGS datasets.}
\vskip -3pt
\centering
\resizebox{\textwidth}{!}{
\begin{tabular}{ c | c c c | c c c }
\hline
\toprule[5.2pt]
 \rowcolor{lightgray}
 {\fontsize{82}{82}\selectfont\textbf{Datasets}} &
 \multicolumn{3}{ c |}{\fontsize{82}{82}\selectfont\textbf{QM9}} & \multicolumn{3}{| c |}{\fontsize{82}{82}\selectfont\textbf{GEOM-DRUGS}} \\
 \midrule[5.2pt]

\fontsize{80}{80}\selectfont Ground-truth &
\includegraphics[width=20cm]{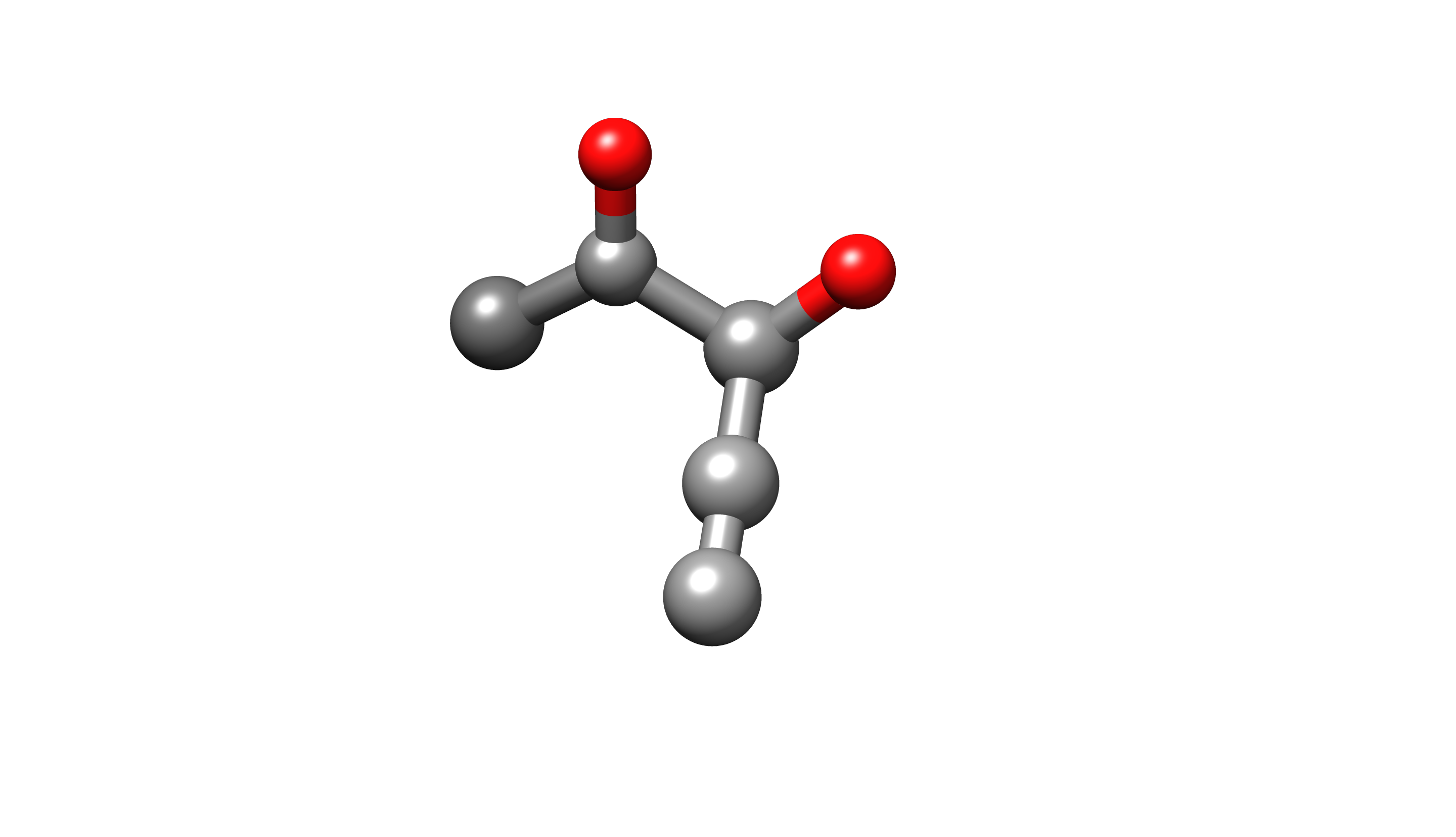} & 
\includegraphics[width=20cm]{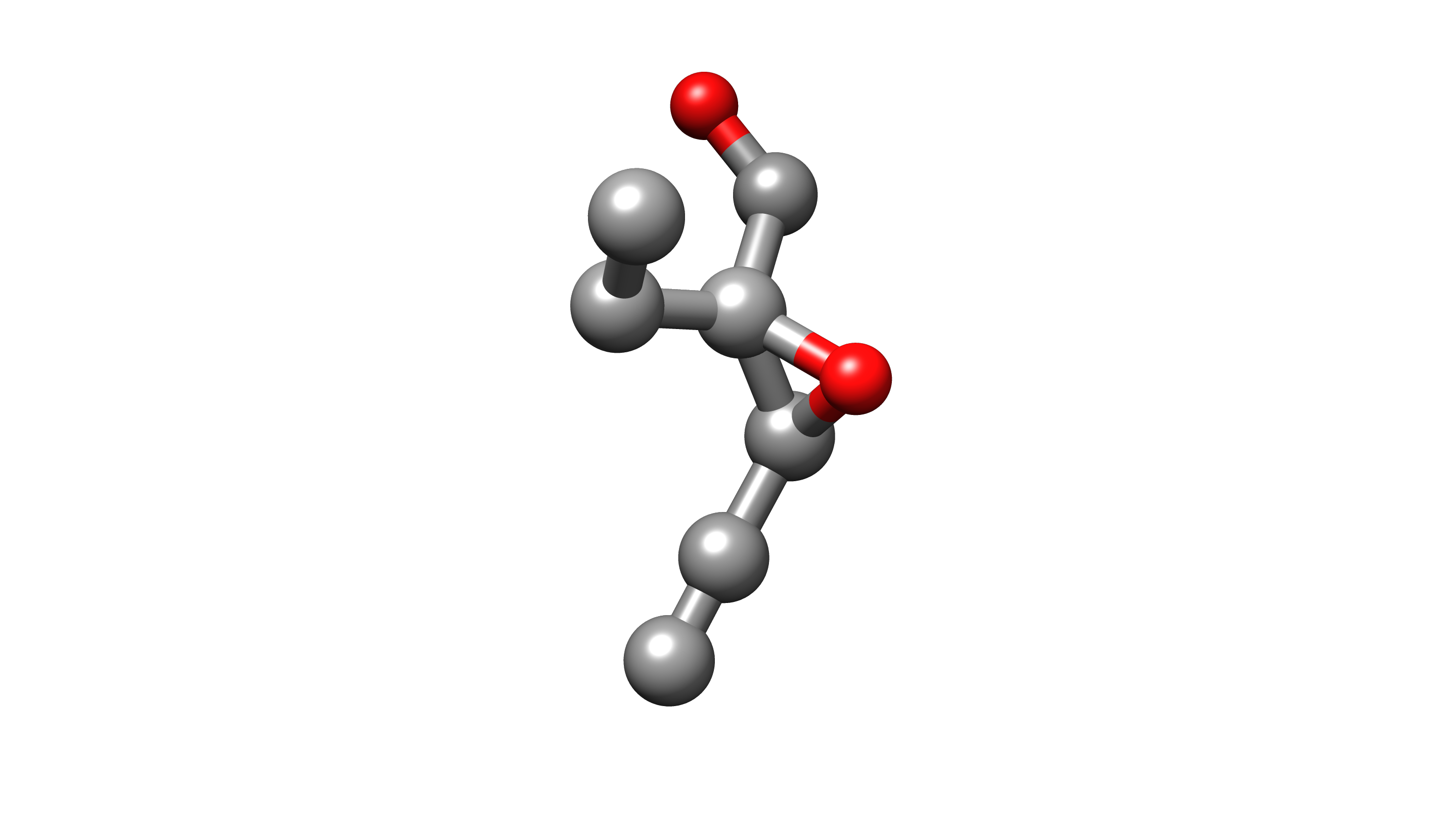} &
\includegraphics[width=20cm]{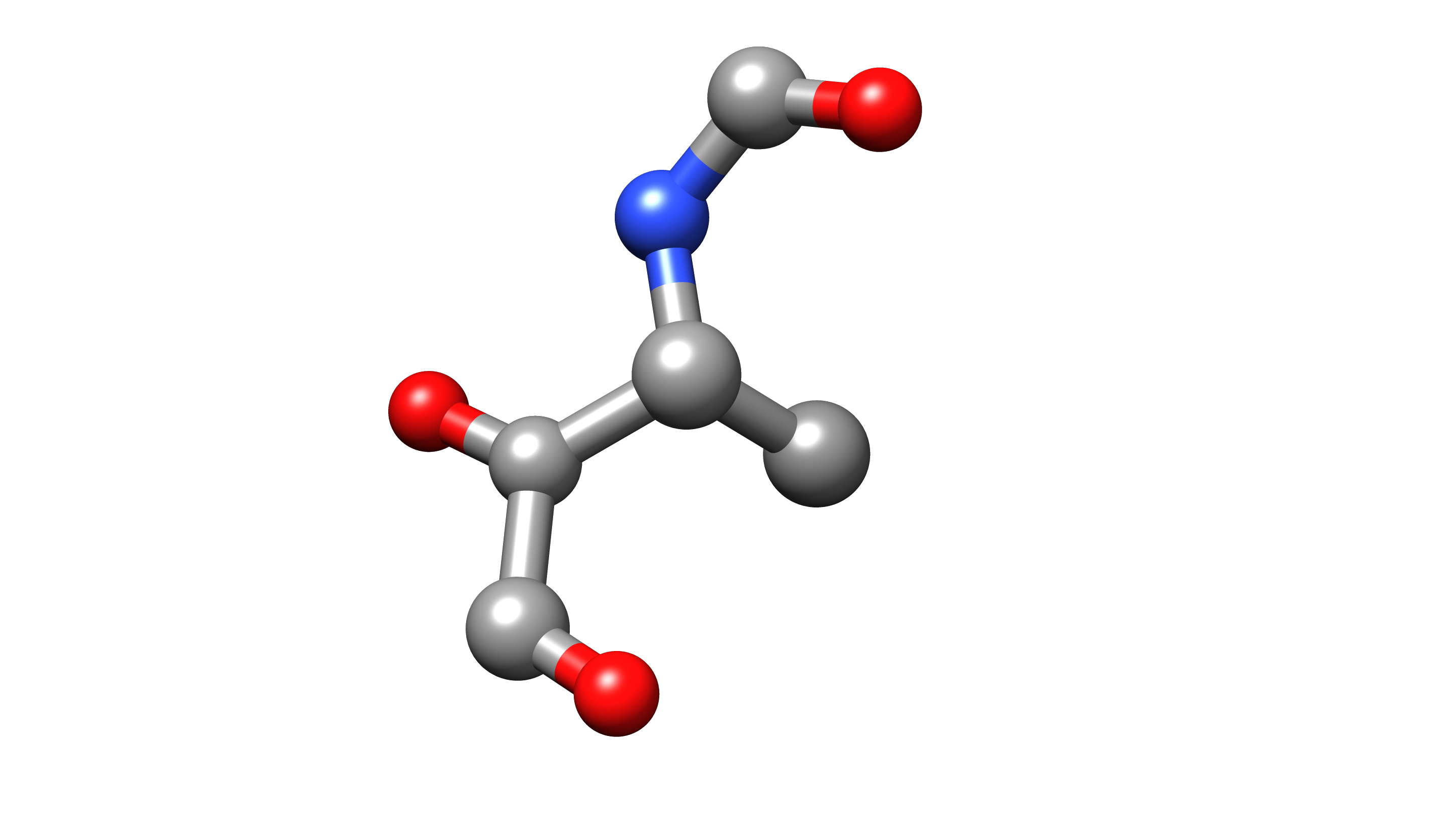} &
\includegraphics[width=20cm]{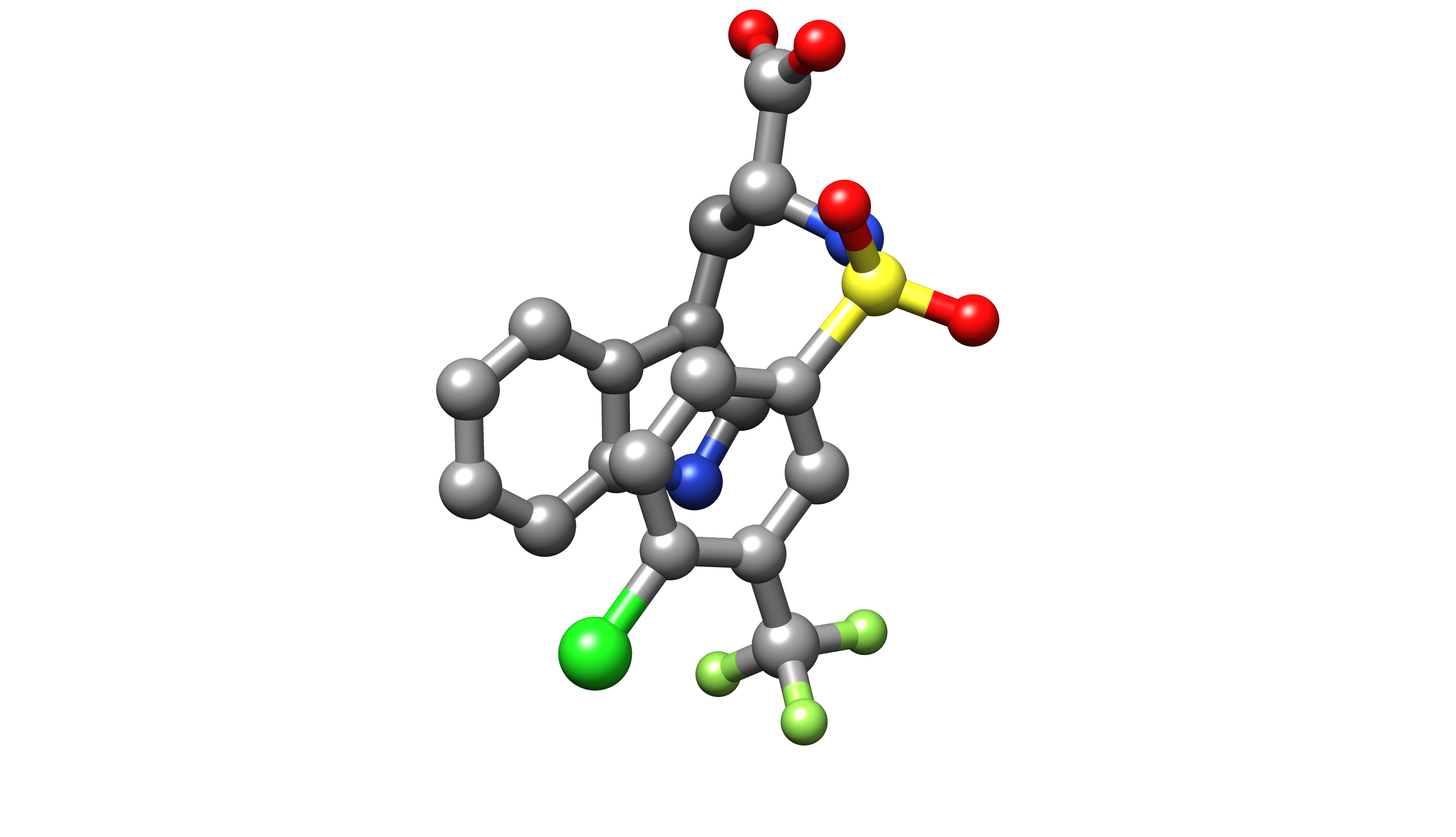} & \includegraphics[width=20cm]{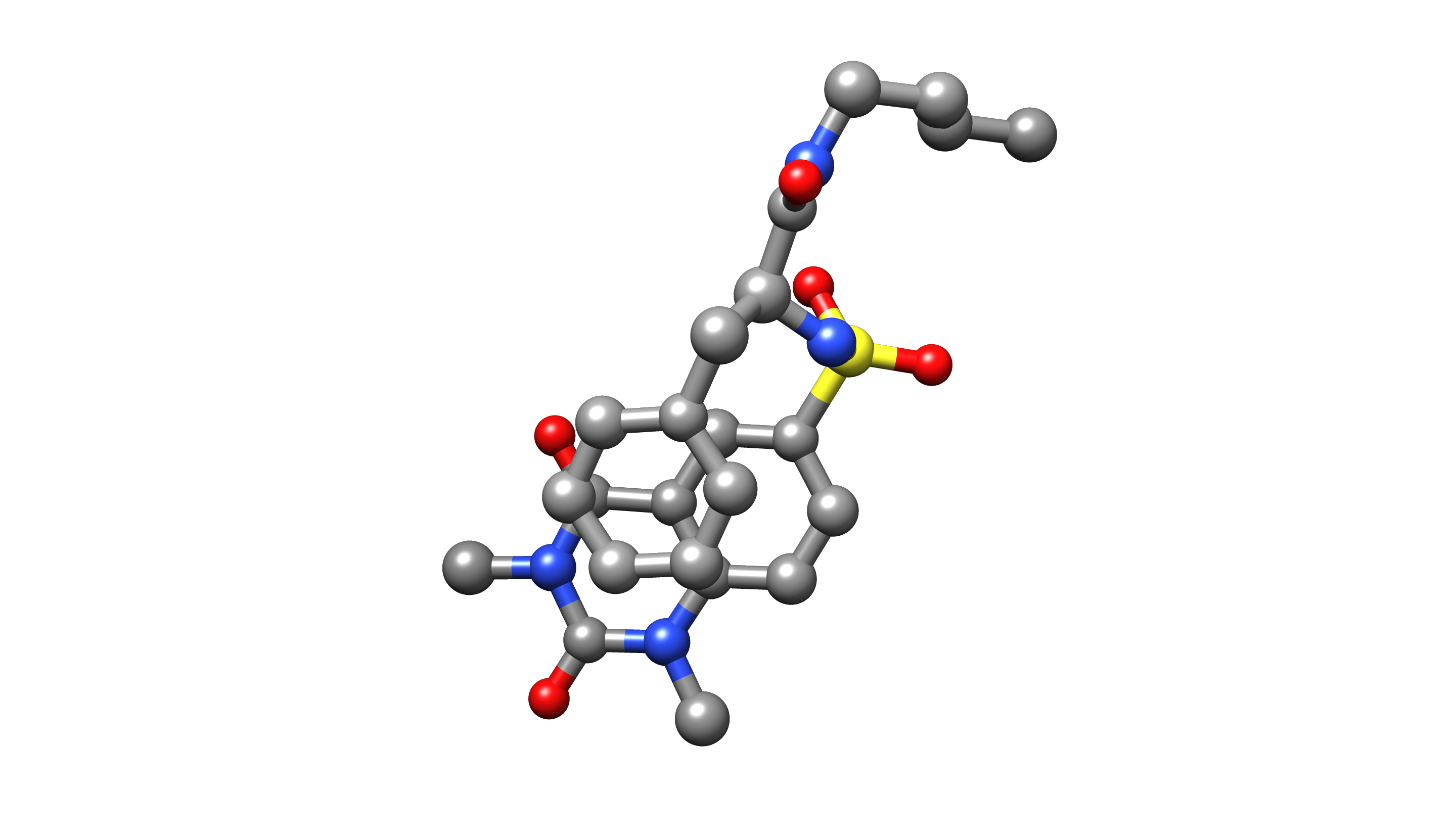} & \includegraphics[width=20cm]{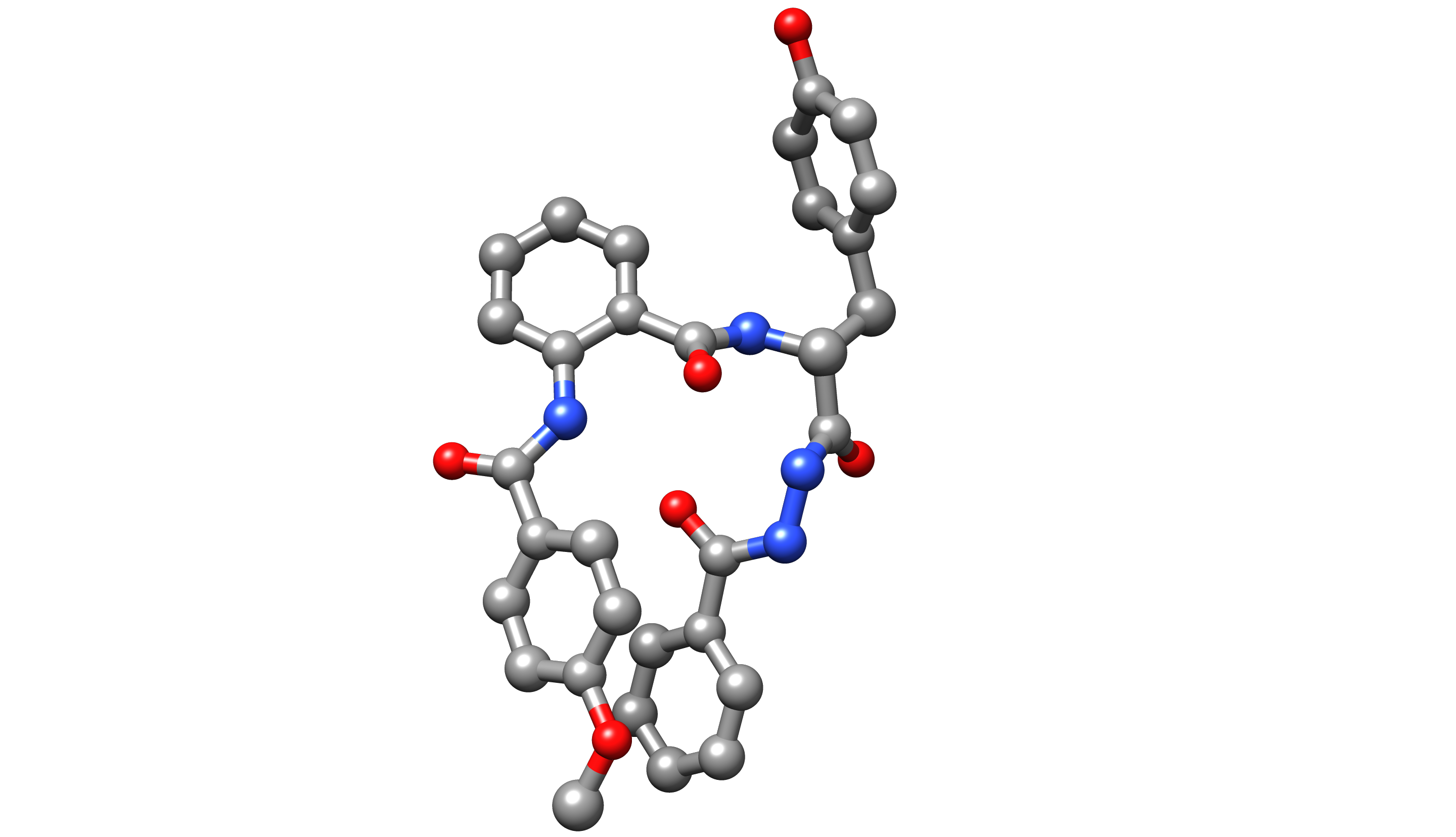} \\
\hline

\fontsize{80}{80}\selectfont RDKit-ETKDG &
\includegraphics[width=20cm]{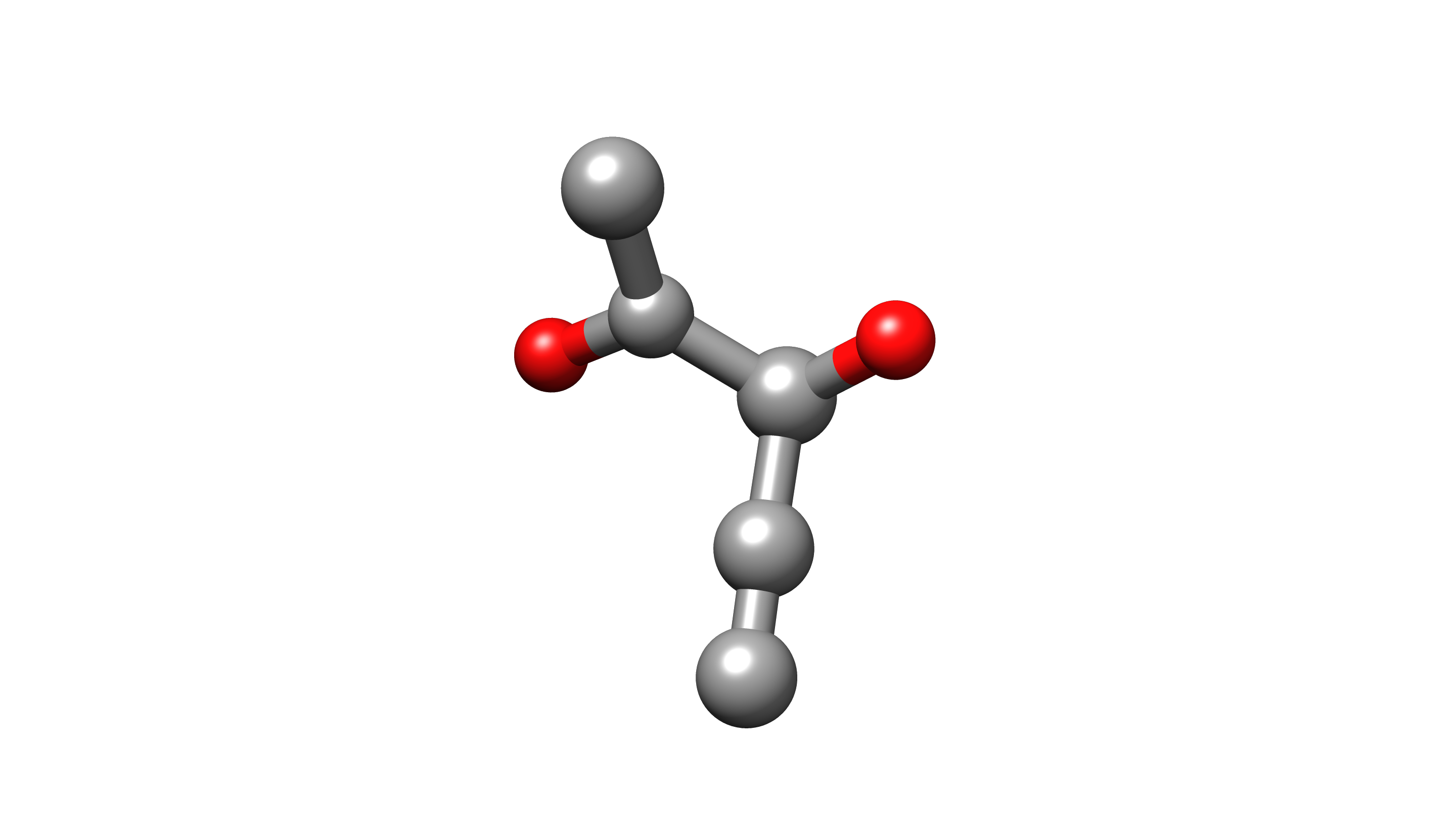} & \includegraphics[width=20cm]{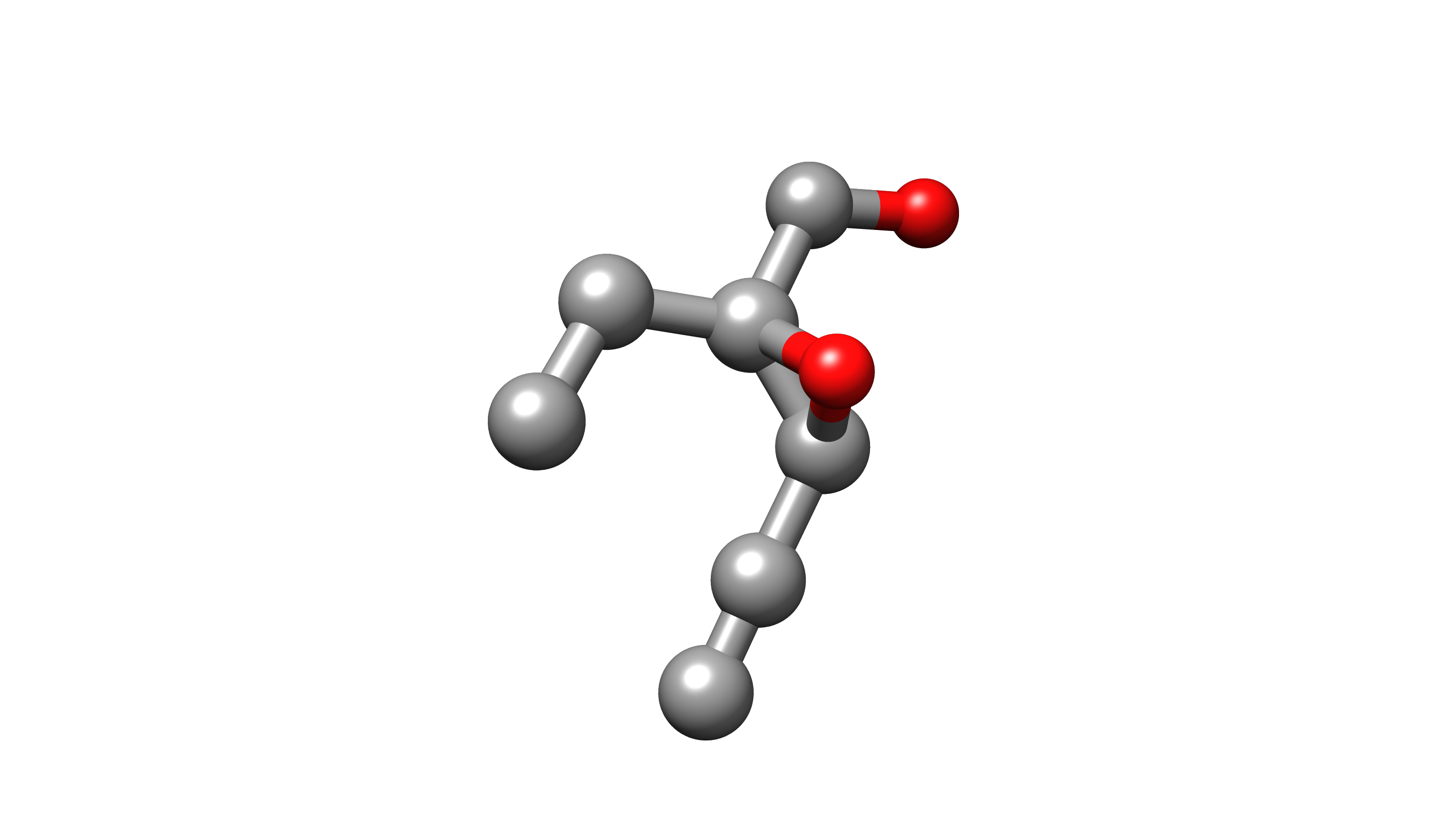} & \includegraphics[width=20cm]{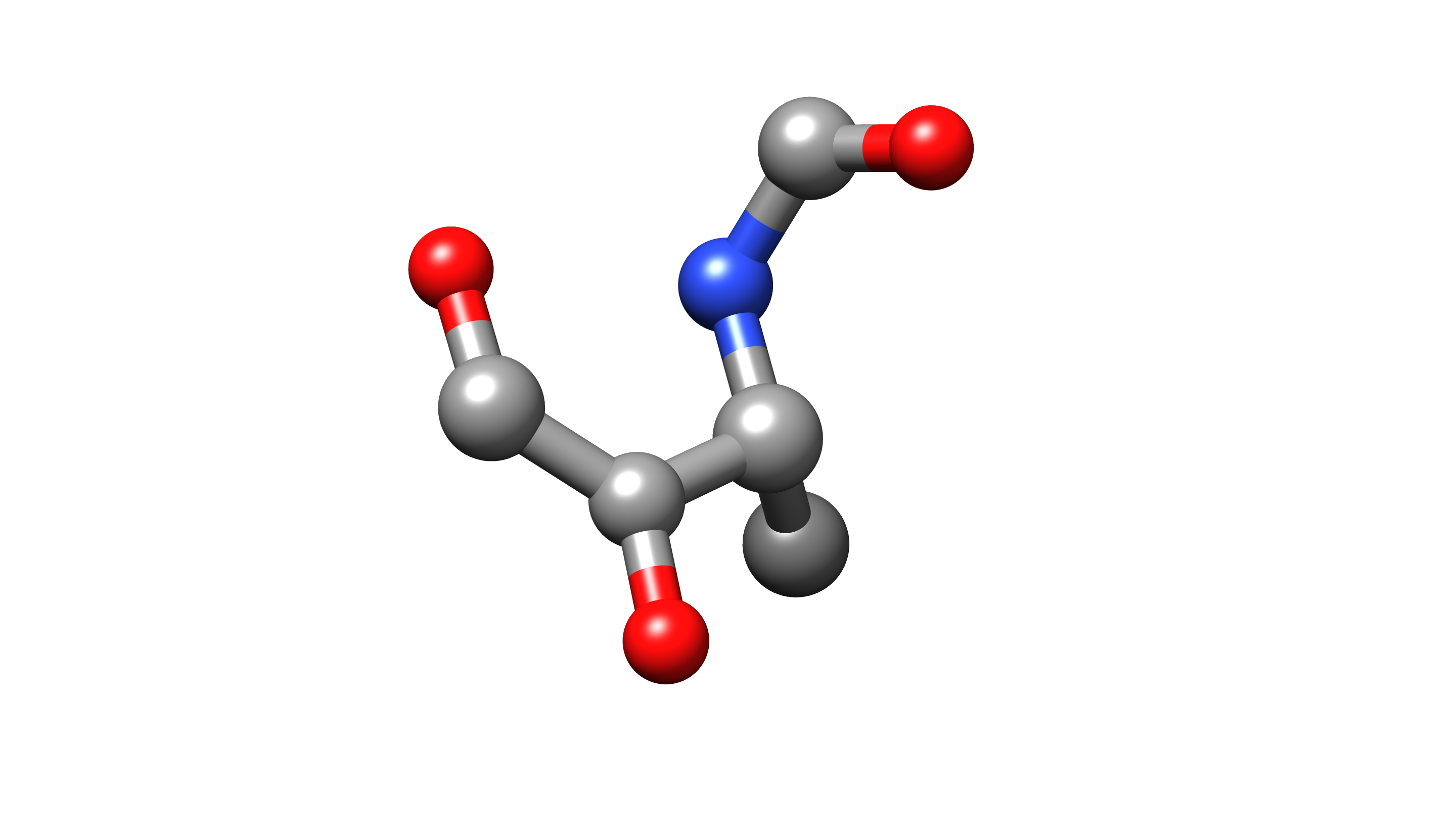} &
\includegraphics[width=20cm]{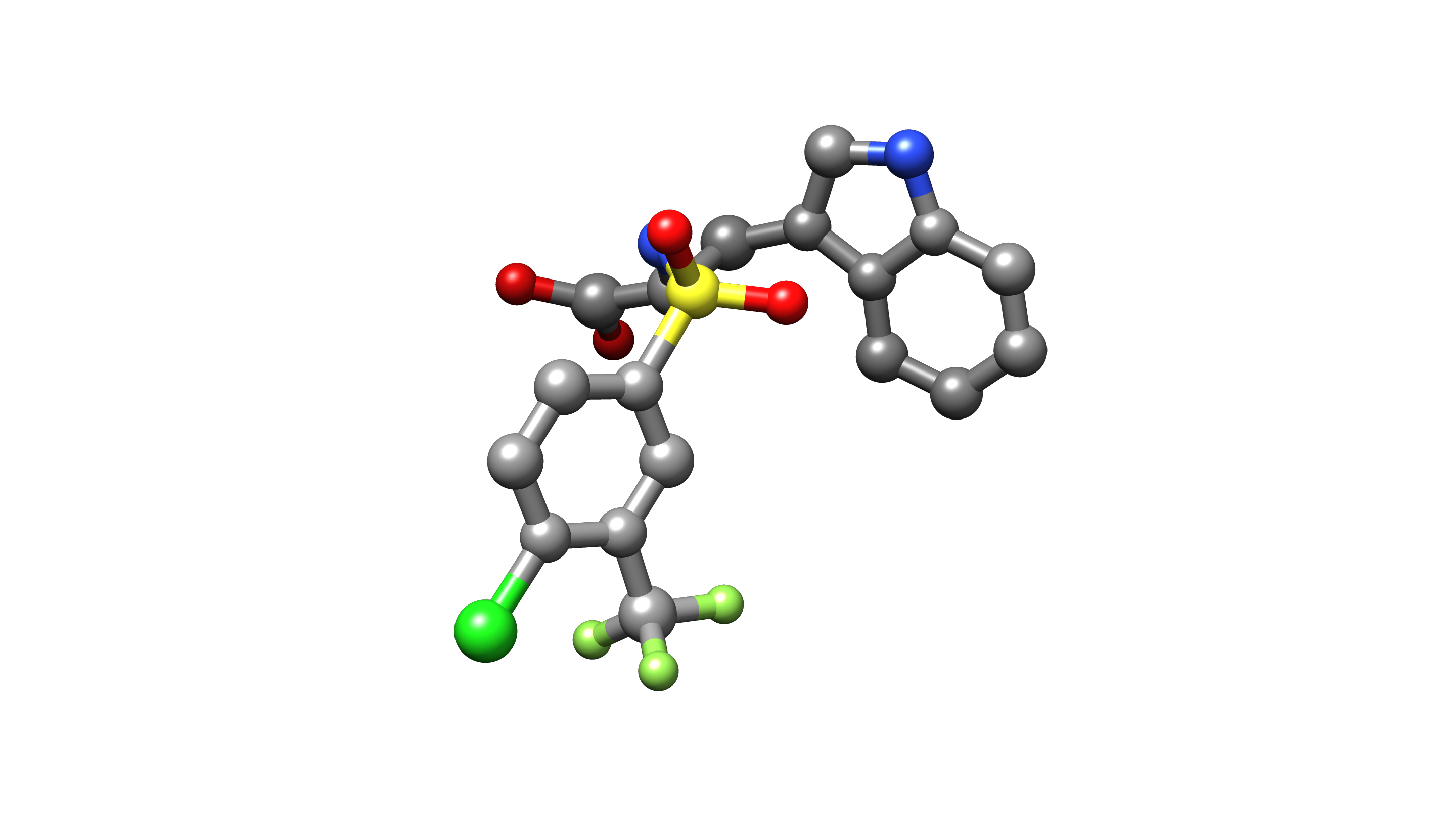} & \includegraphics[width=20cm]{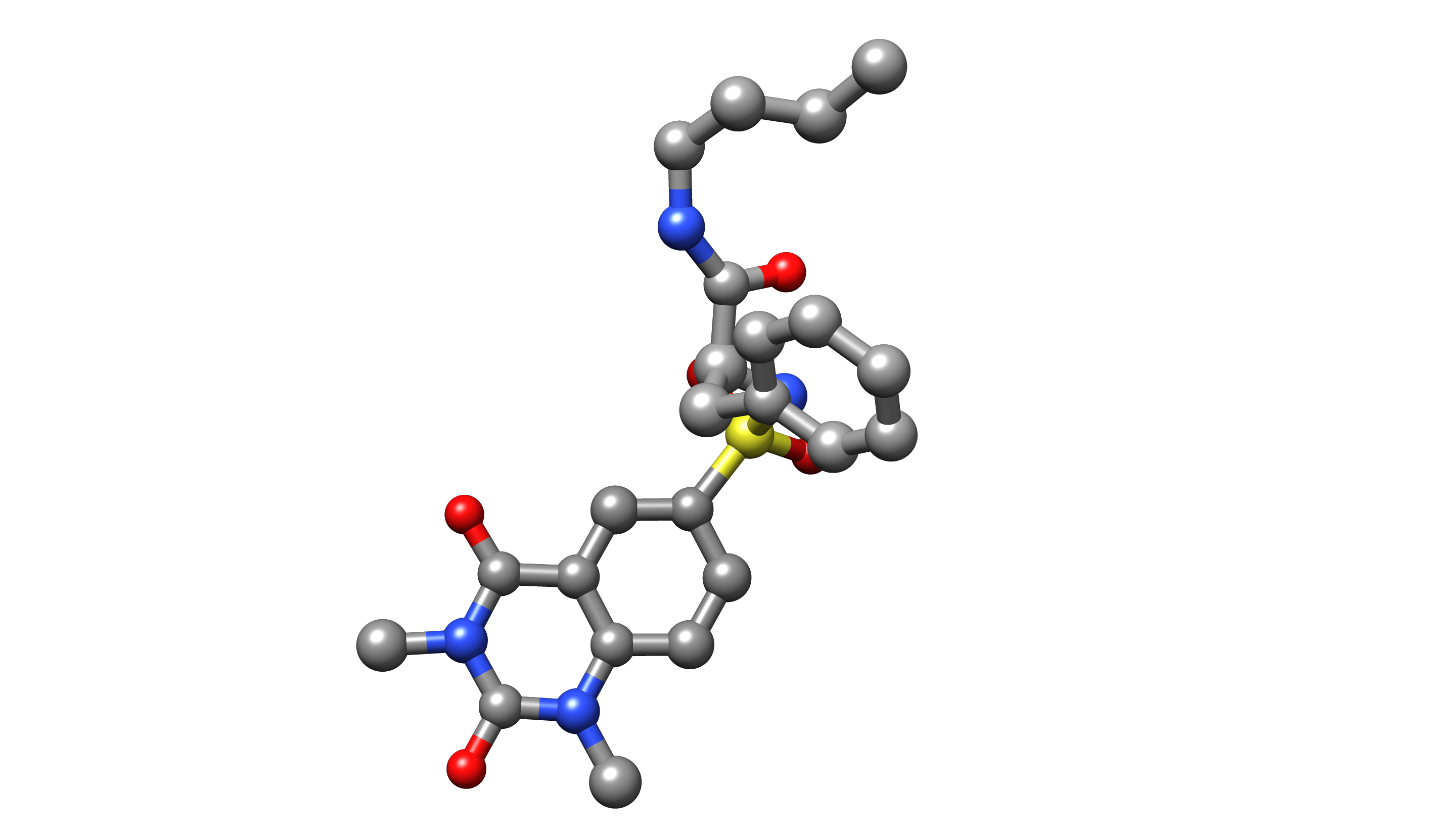} & \includegraphics[width=20cm]{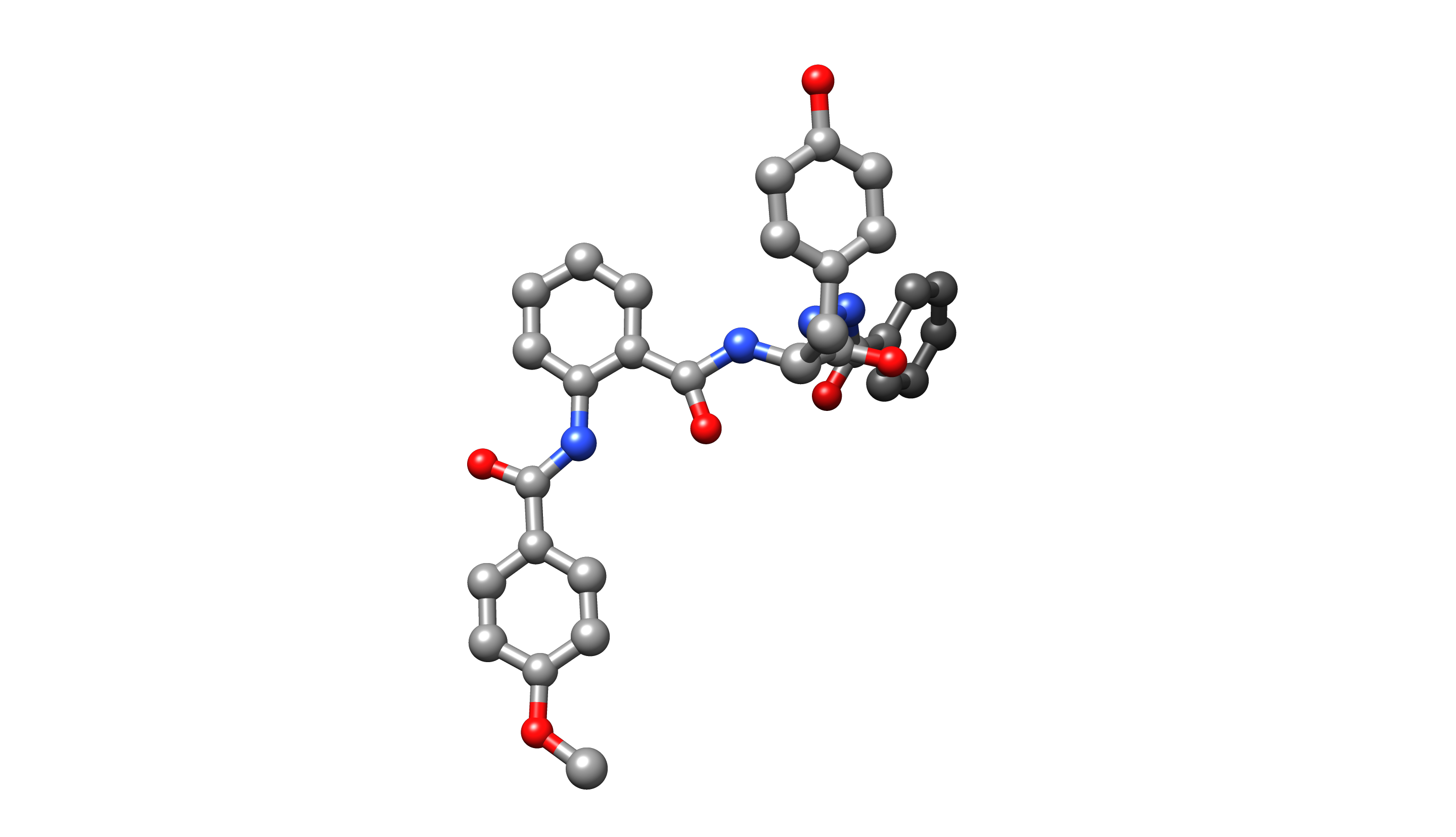} \\
& \fontsize{60}{60}\selectfont 1.233 & 
\fontsize{60}{60}\selectfont 0.844 & 
\fontsize{60}{60}\selectfont 1.621 & 
\fontsize{60}{60}\selectfont 3.033 & 
\fontsize{60}{60}\selectfont 2.515 & 
\fontsize{60}{60}\selectfont 3.470 \\
\hline

\fontsize{80}{80}\selectfont GTMGC &
\includegraphics[width=20cm]{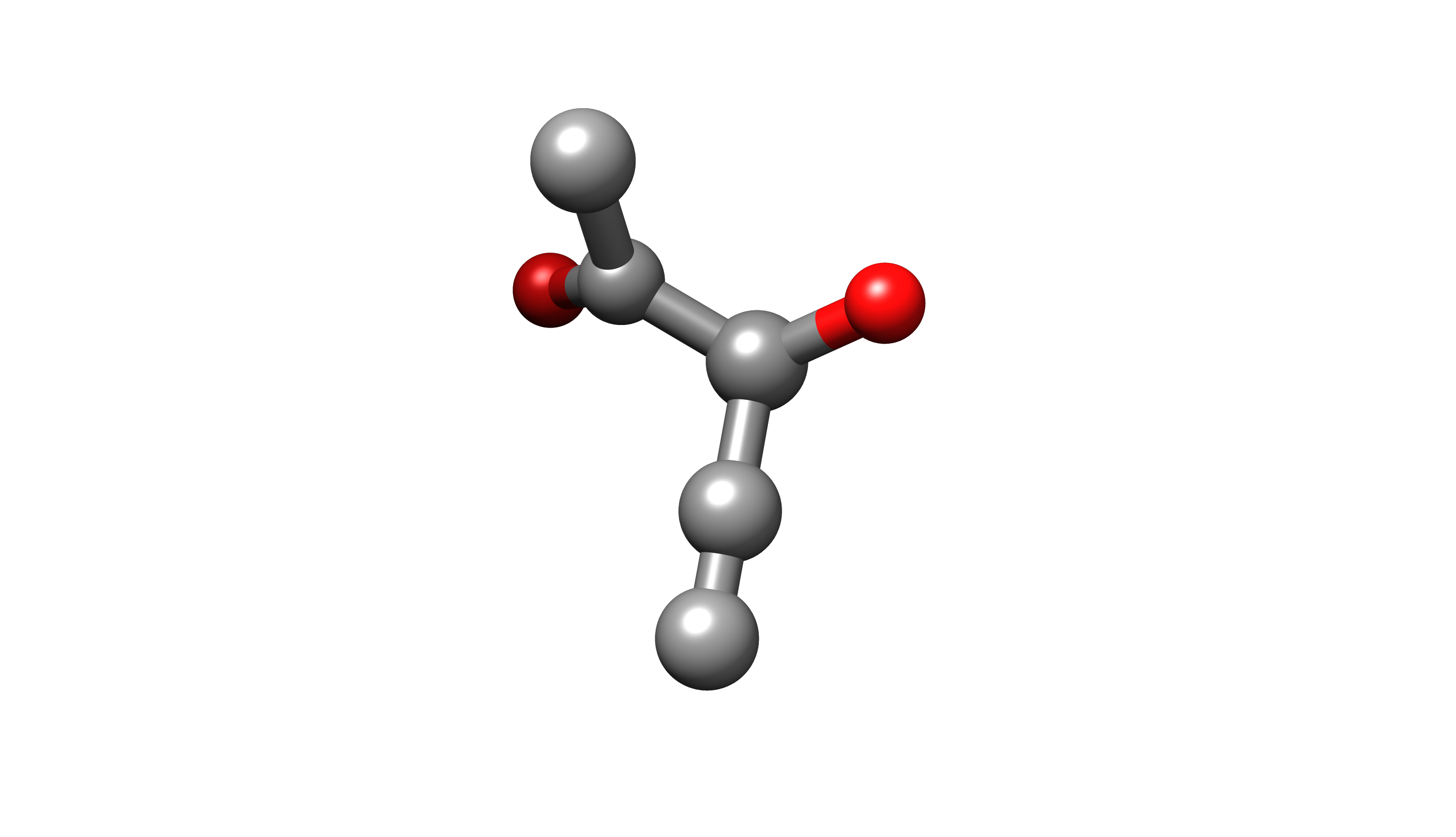} & \includegraphics[width=20cm]{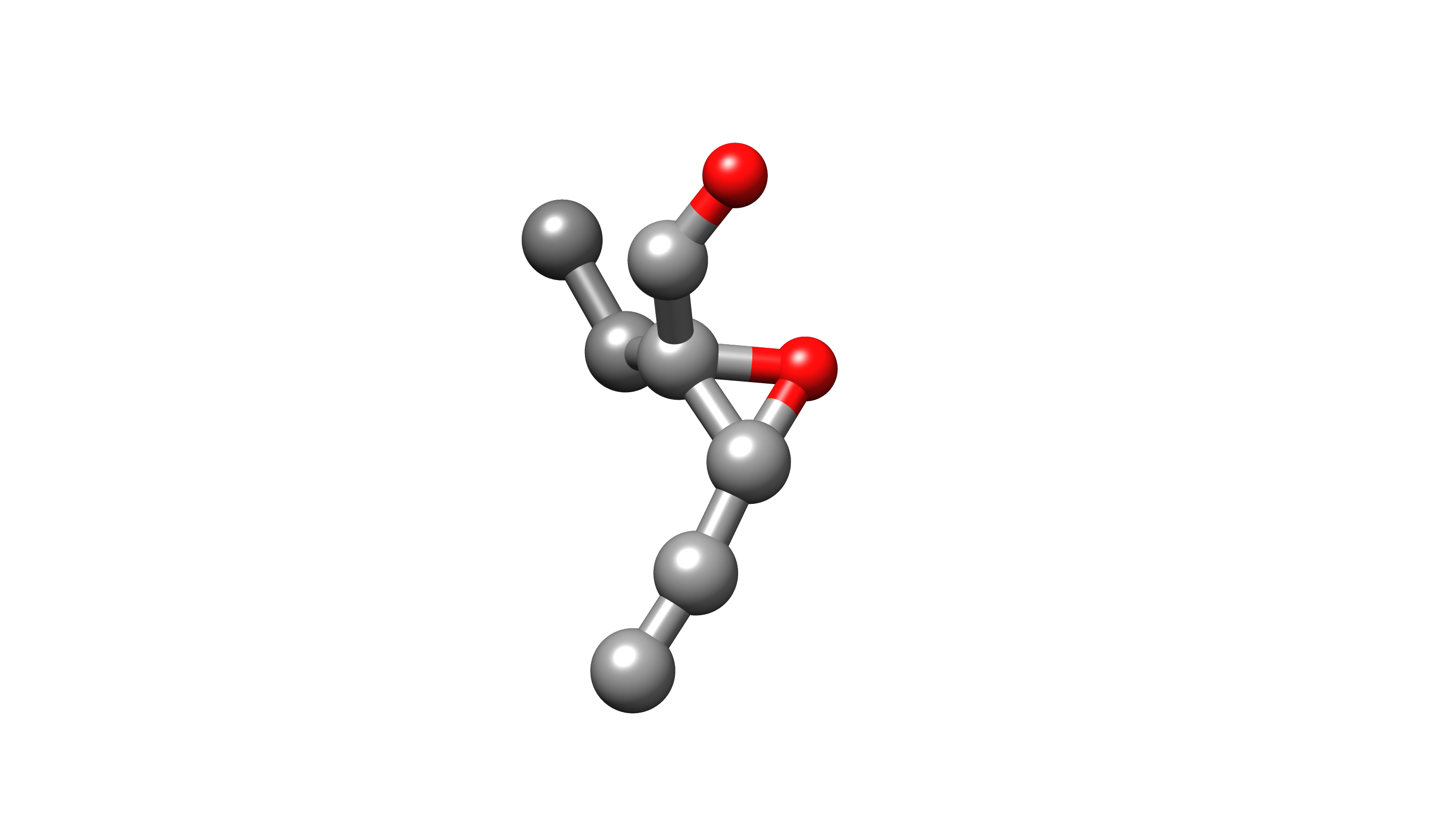} & \includegraphics[width=20cm]{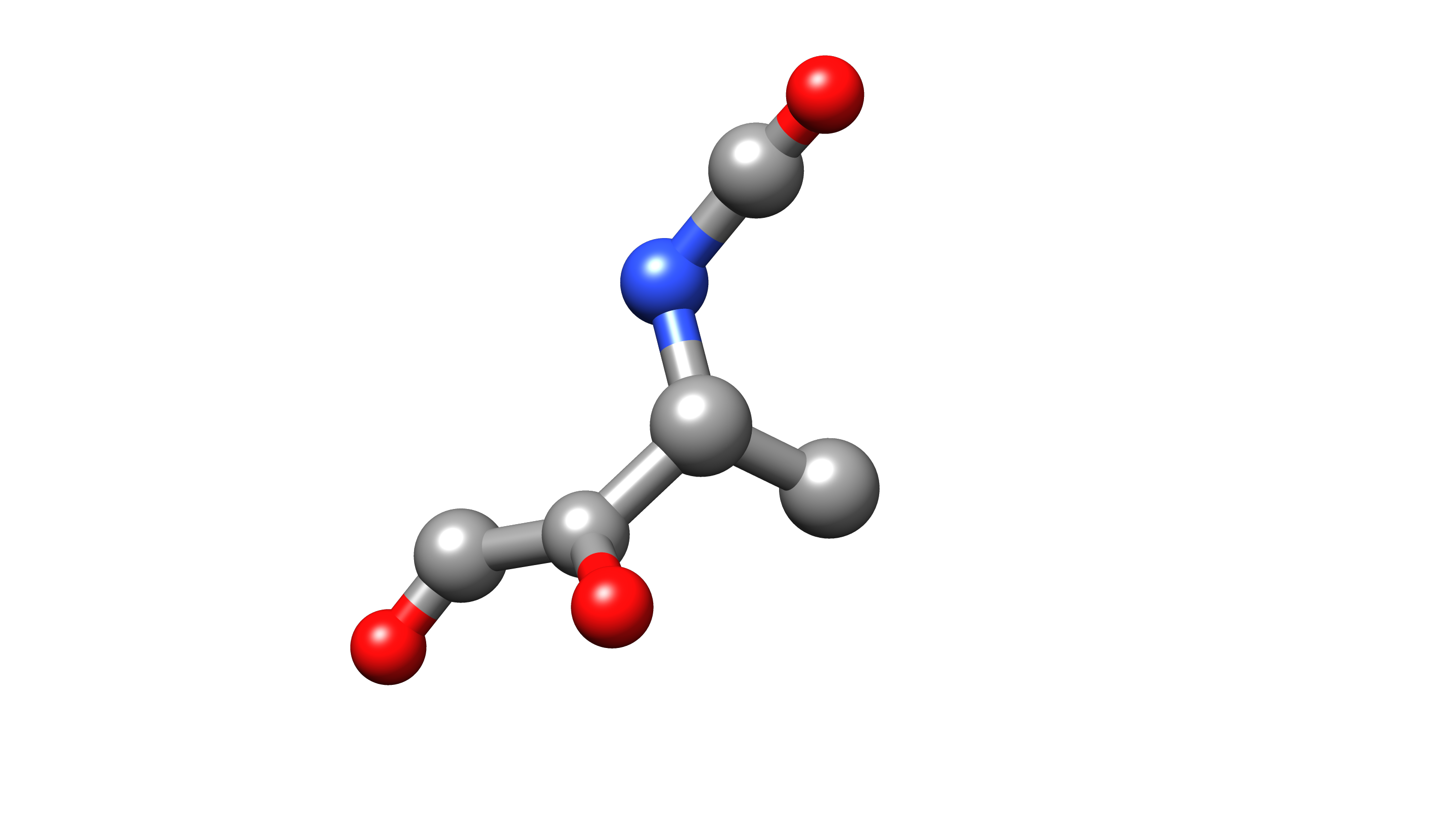} &
\includegraphics[width=20cm]{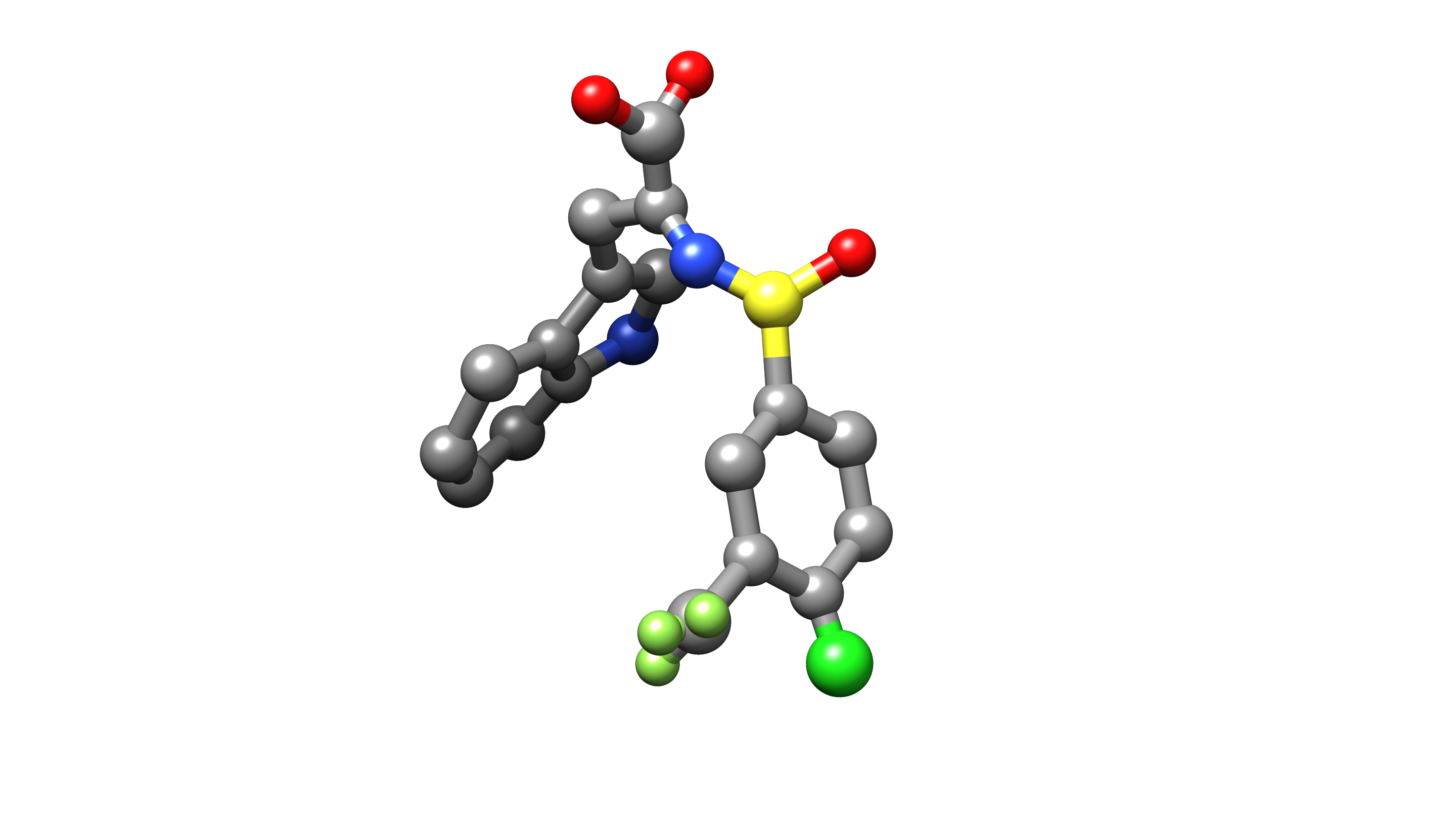} & \includegraphics[width=20cm]{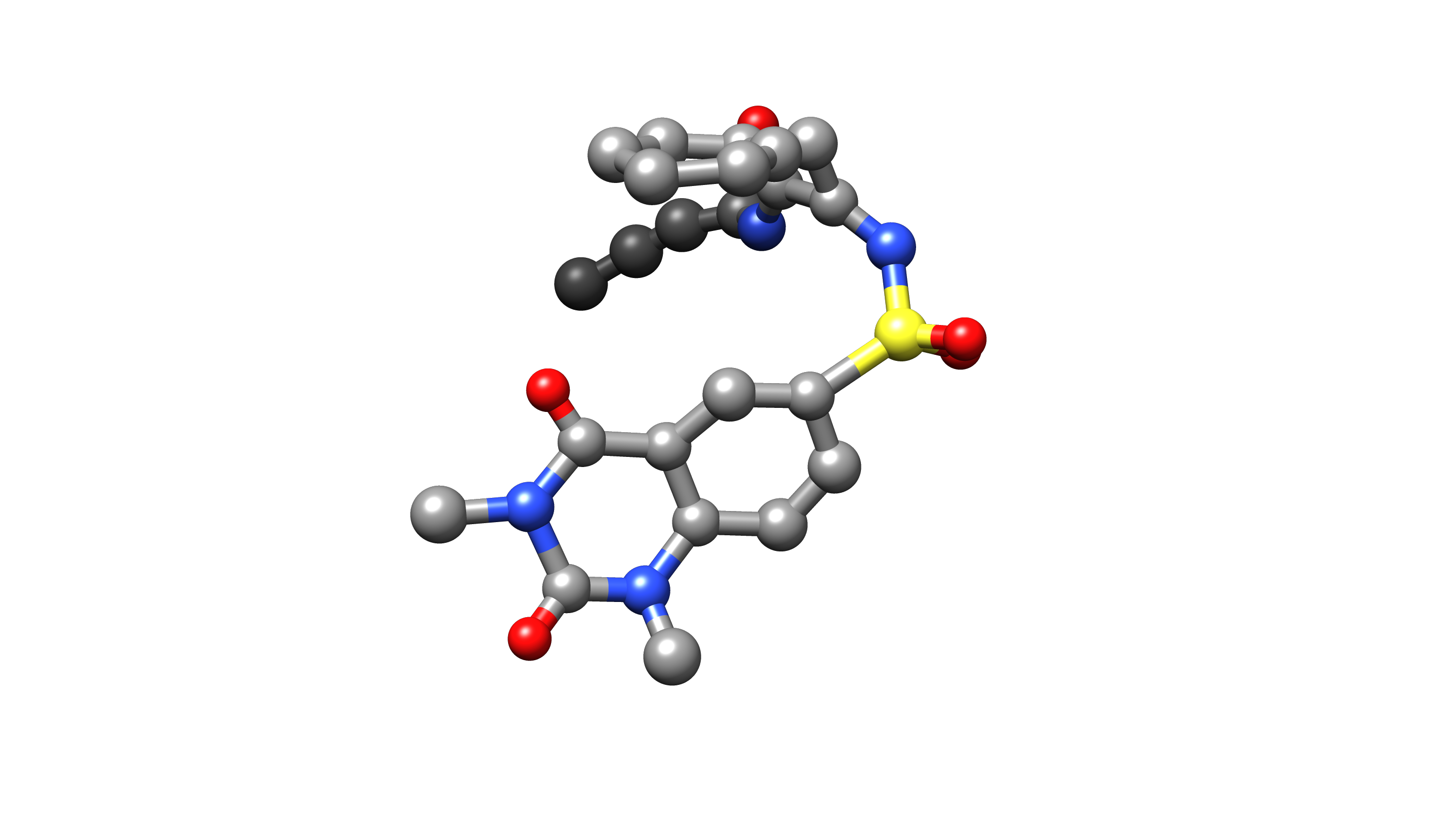} & \includegraphics[width=20cm]{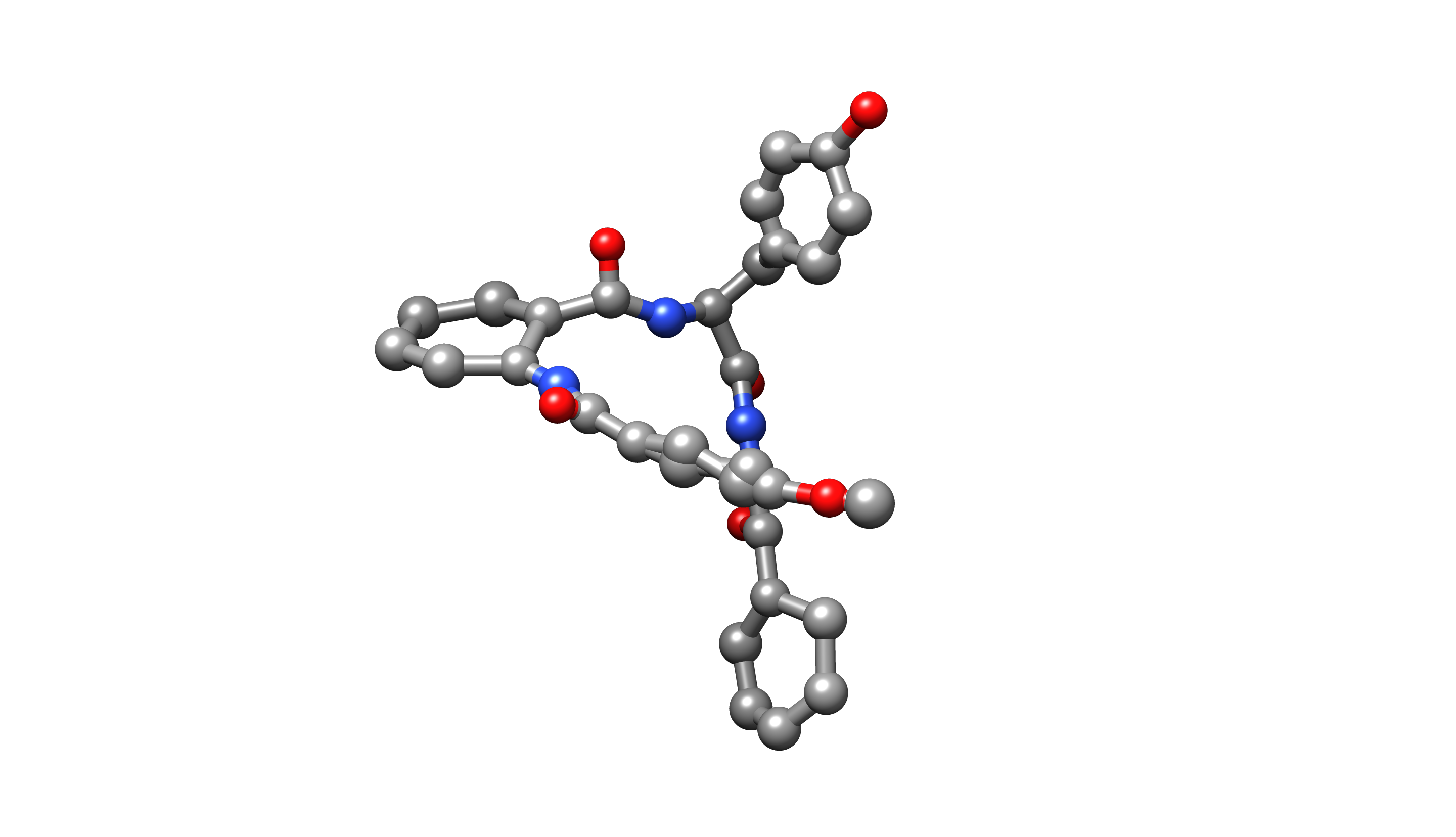} \\
& \fontsize{60}{60}\selectfont 1.242 & 
\fontsize{60}{60}\selectfont 1.168 & 
\fontsize{60}{60}\selectfont 1.479 & 
\fontsize{60}{60}\selectfont 2.295 & 
\fontsize{60}{60}\selectfont 2.232 & 
\fontsize{60}{60}\selectfont 2.521 \\
\hline

\fontsize{80}{80}\selectfont \system{} &
\includegraphics[width=20cm]{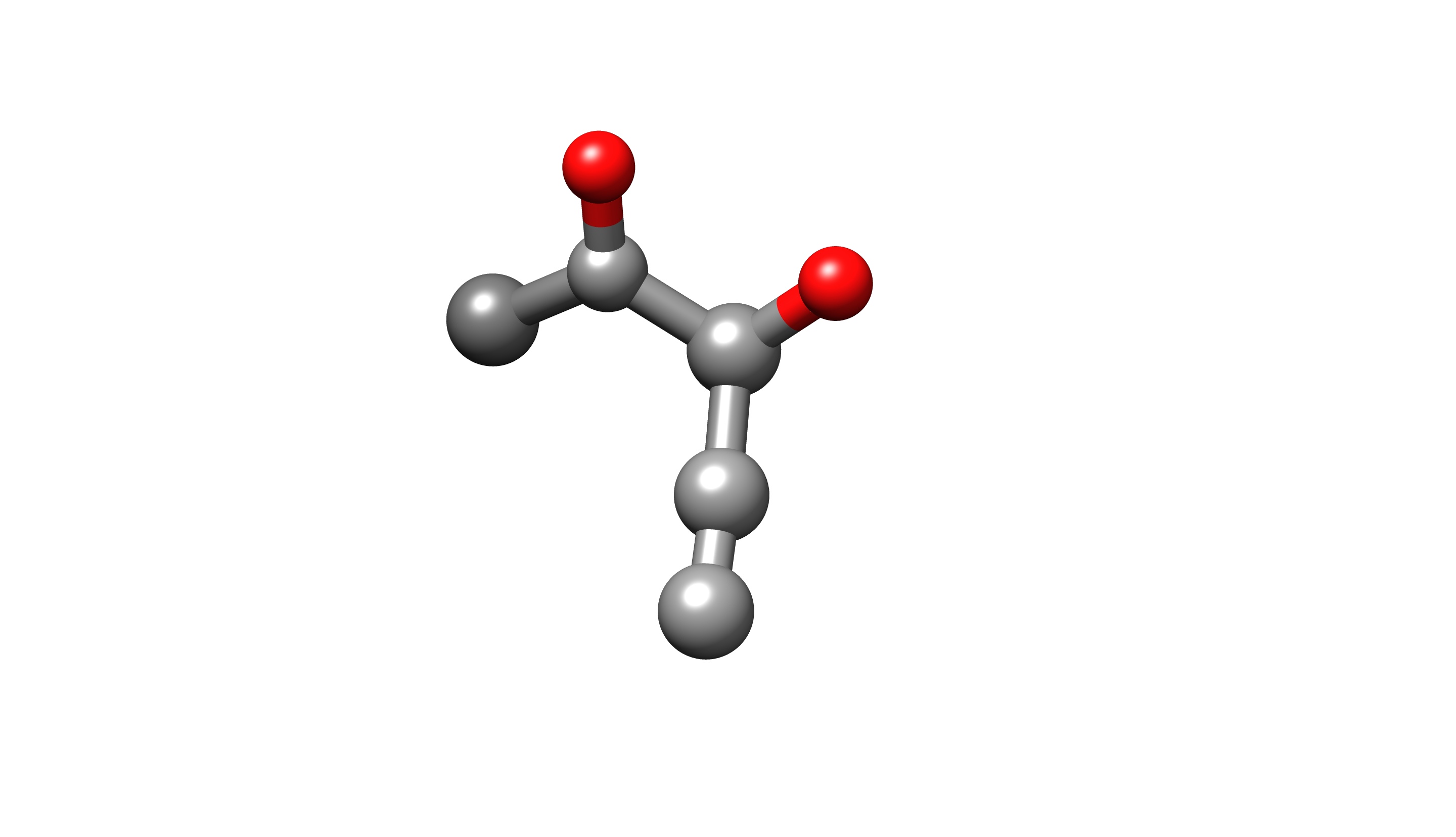} & \includegraphics[width=20cm]{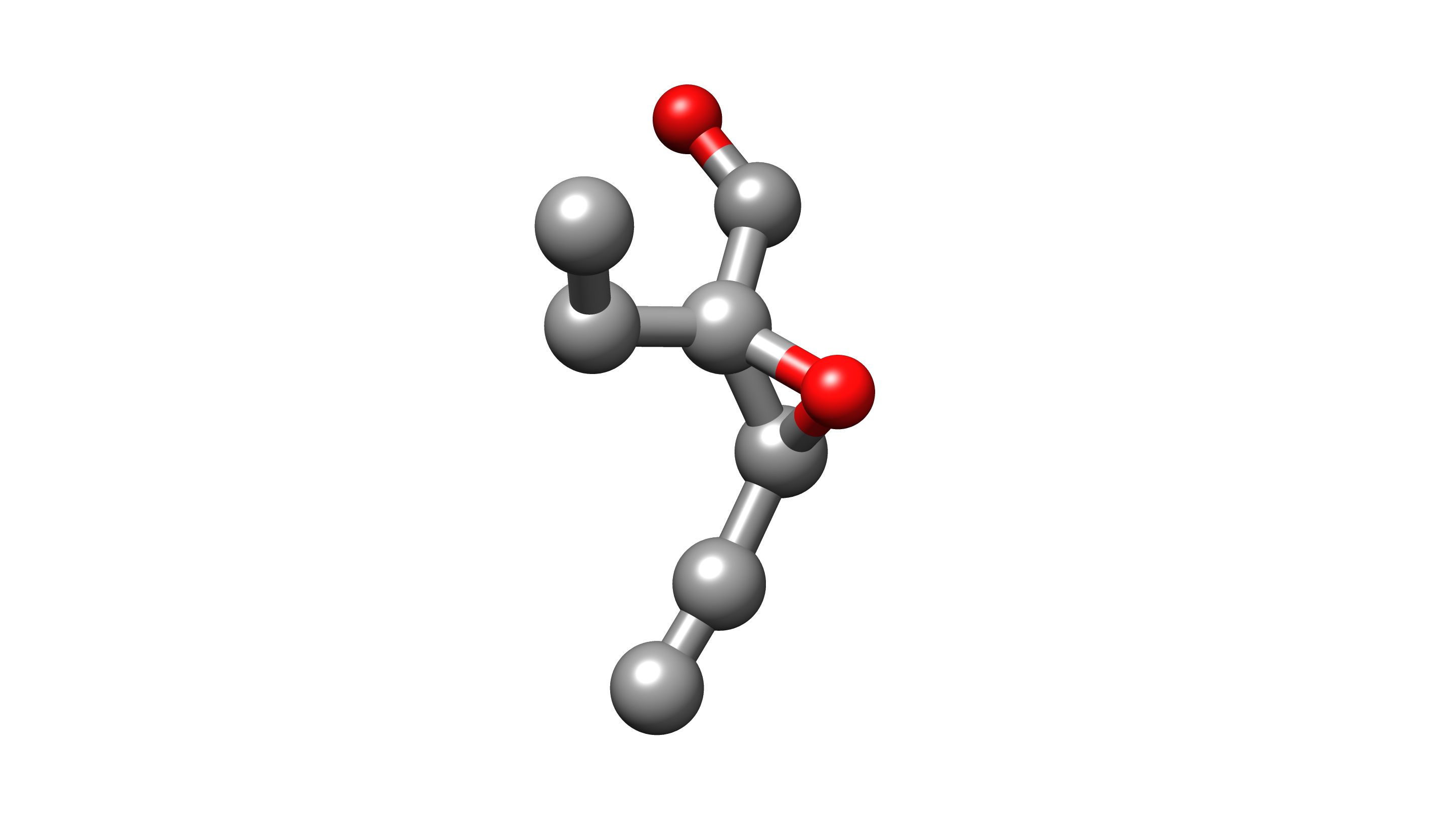} & \includegraphics[width=20cm]{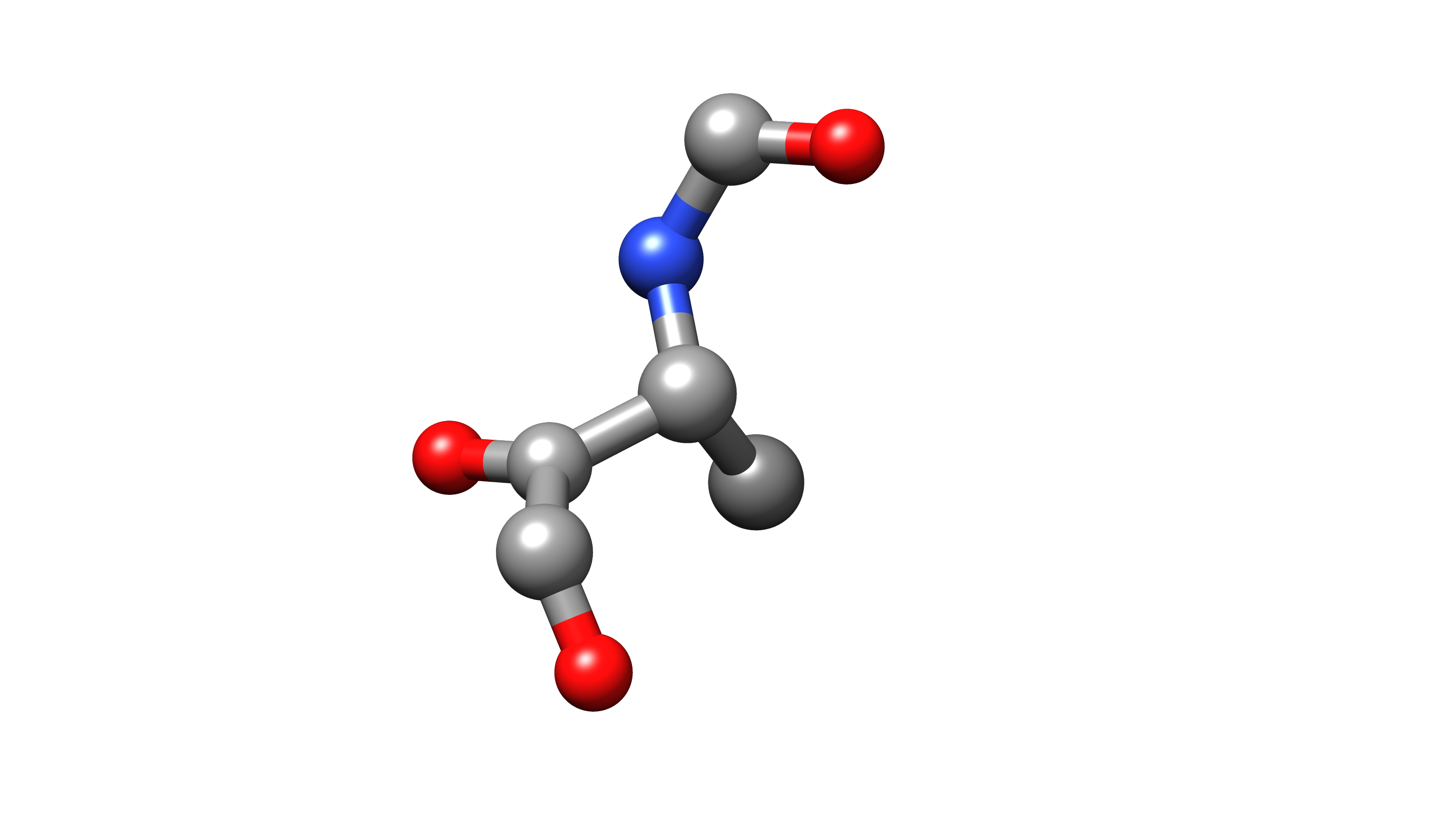} &
\includegraphics[width=20cm]{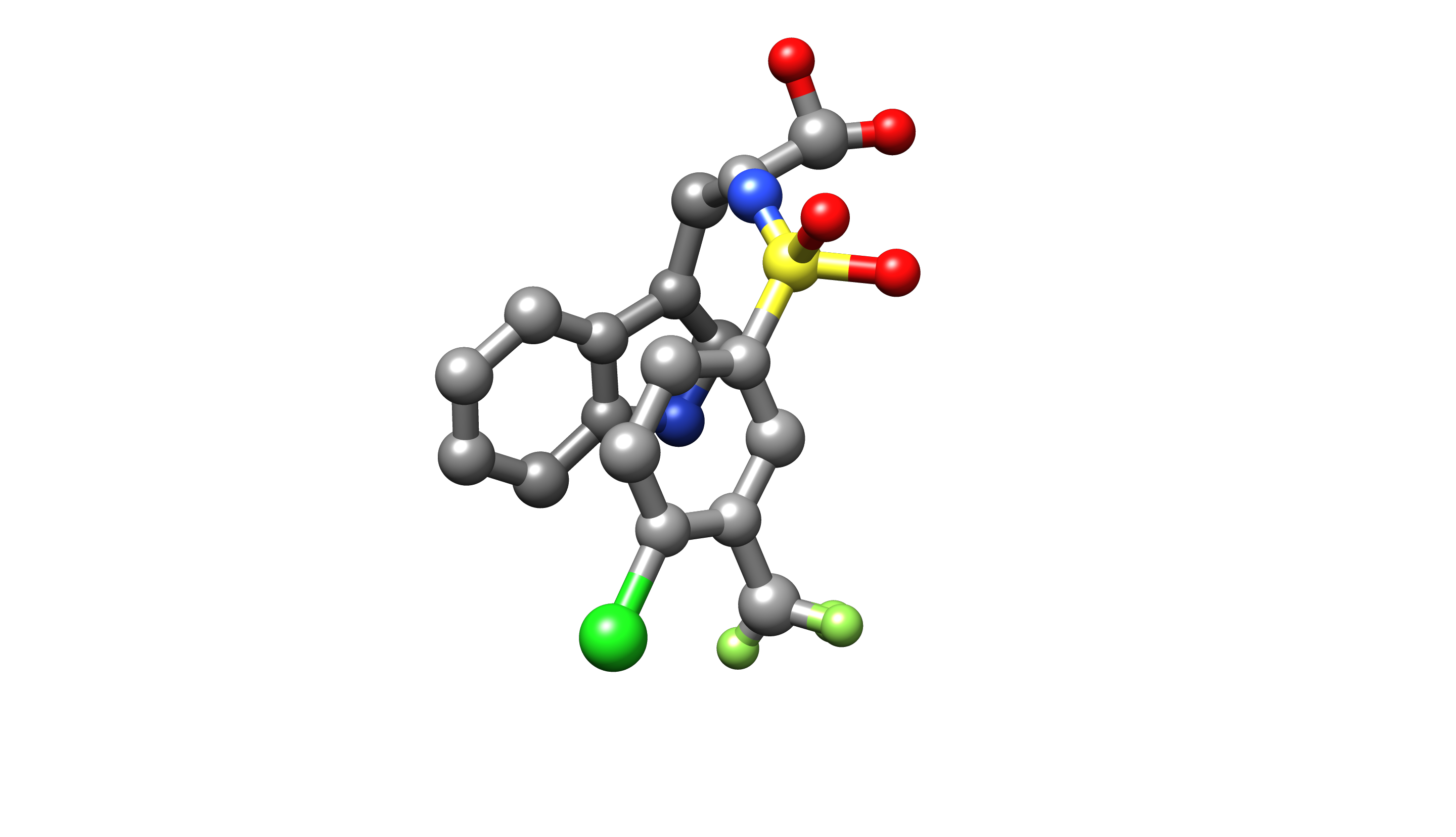} & \includegraphics[width=20cm]{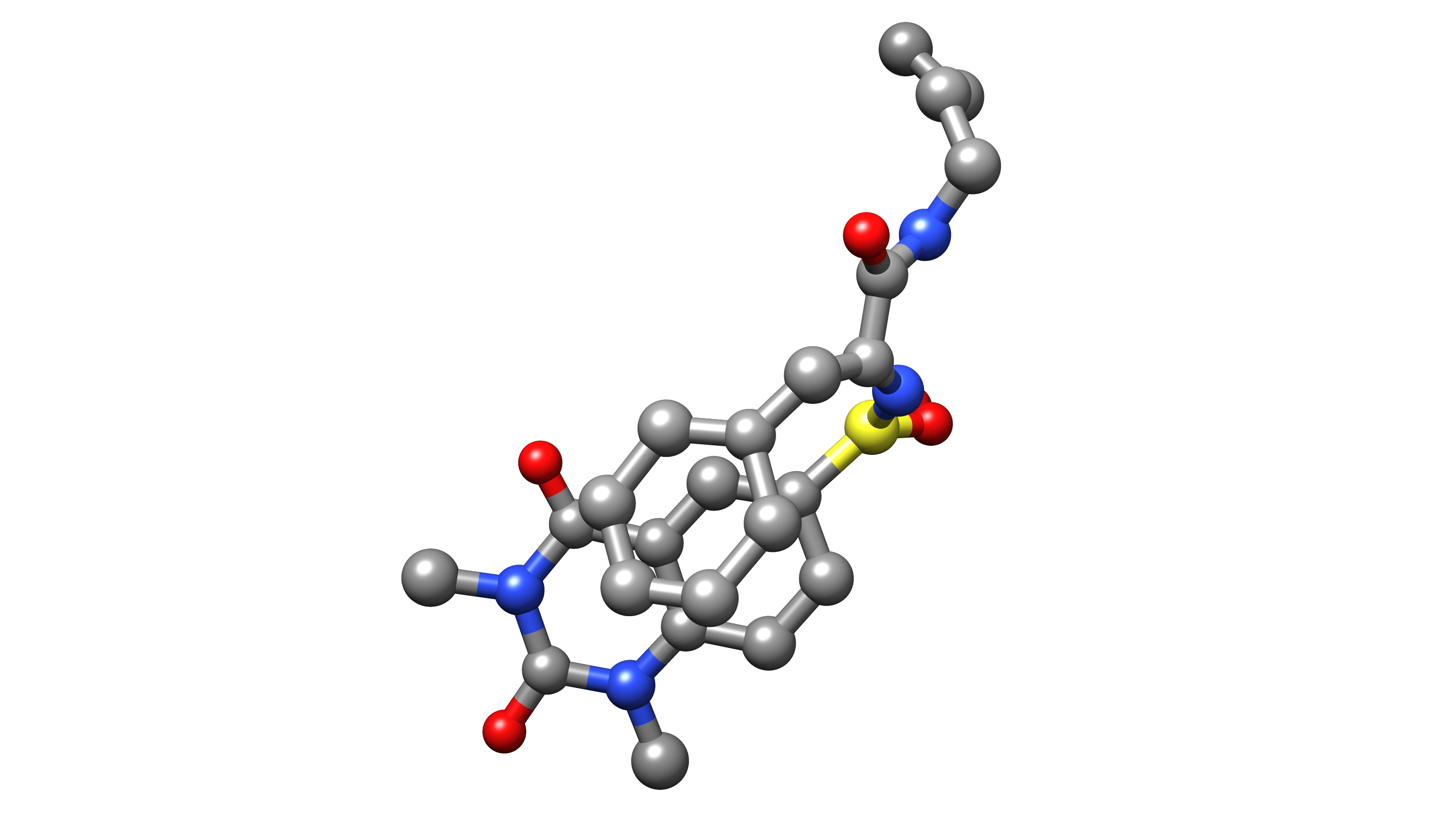} & \includegraphics[width=20cm]{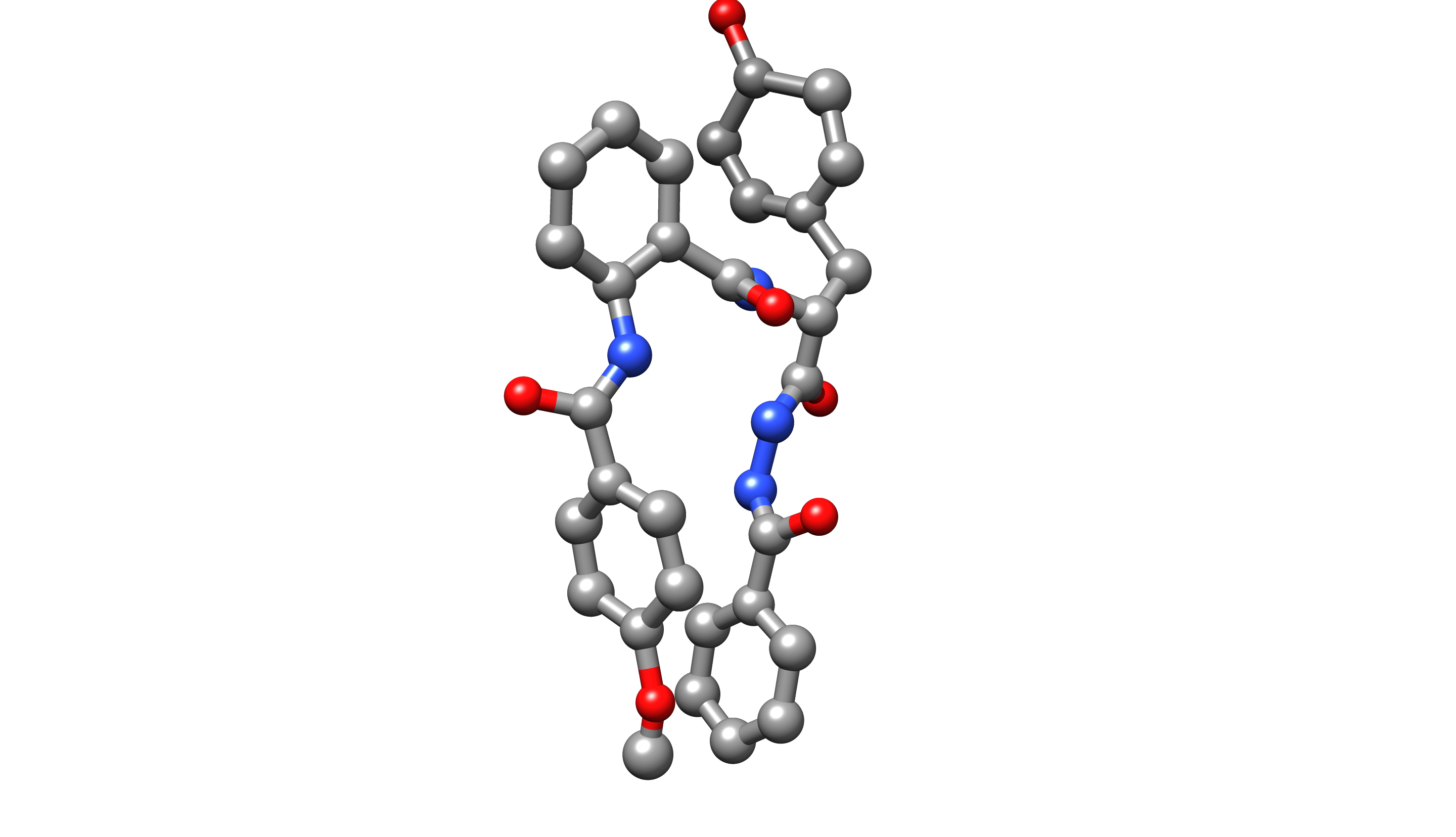} \\
& \fontsize{60}{60}\selectfont \textbf{0.050} & 
\fontsize{60}{60}\selectfont \textbf{0.058} & 
\fontsize{60}{60}\selectfont \textbf{0.376} & 
\fontsize{60}{60}\selectfont \textbf{0.844} & 
\fontsize{60}{60}\selectfont \textbf{0.898} & 
\fontsize{60}{60}\selectfont \textbf{1.174} \\
\bottomrule[5.2pt]
\end{tabular}
}
\vskip -10 pt
\end{table*}
\section{Additional Experiments}\label{B}

\subsection{Molecular Visualizations}
In this section, we provide 3D visulaizations and corresponding RMSD predictions from our method, RDKit-ETKDG, and GTMGC on the QM9 and GEOM-DRUGS dataset. As illustrated, the conformations predicted by \system{} shows significantly closer alignment with the ground-truth conformations. This superiority is consistent across molecules of varying sizes, demonstrating the robustness of our approach.

\begin{wrapfigure}{r}{0.5\linewidth}
    \vskip -13pt
    \includegraphics[width=\linewidth]{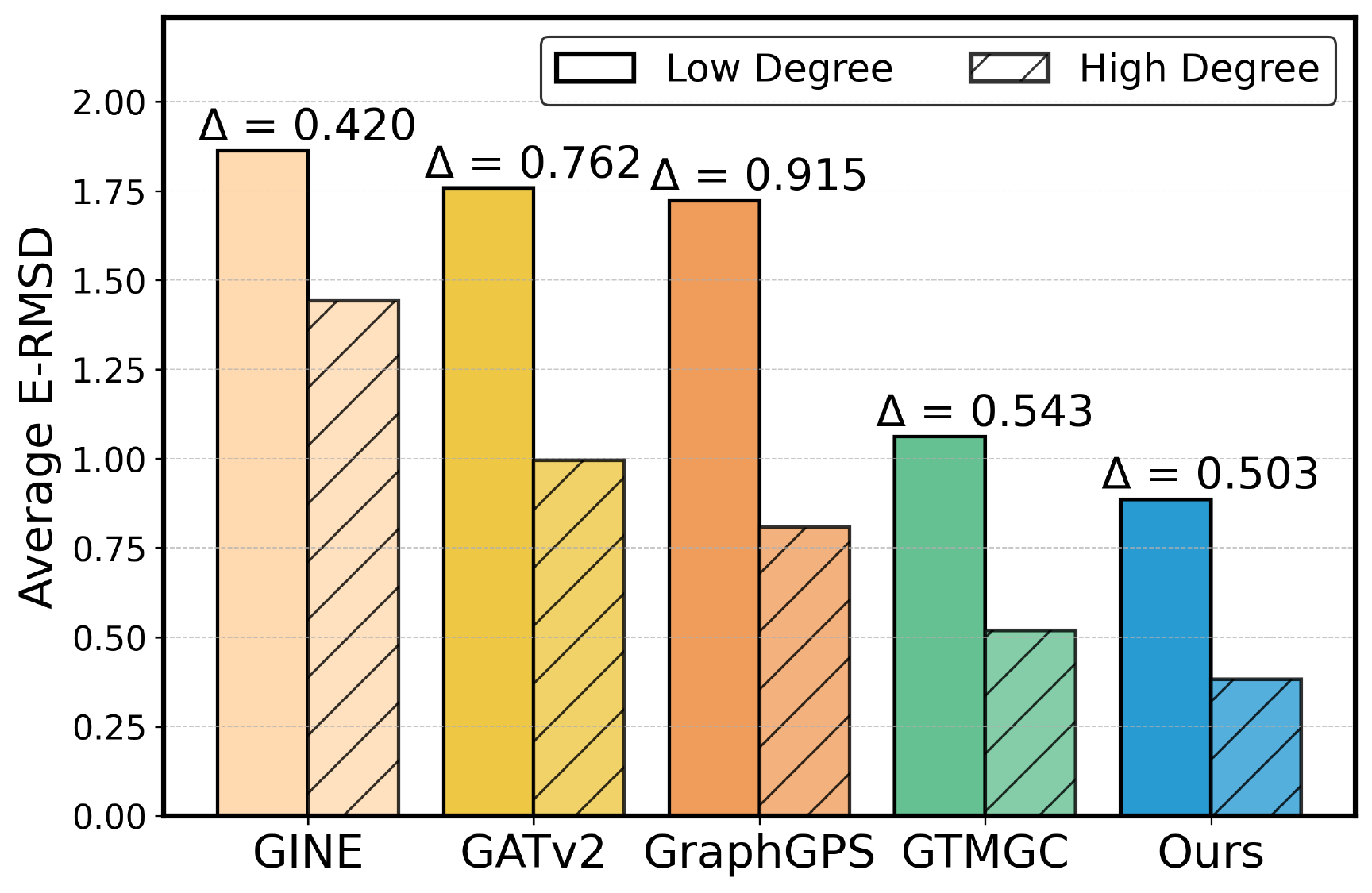} 
    \vskip -5pt
    \caption{Atom-wise E-RMSD analysis with respect to the relative atom degree on \system{} and baselines. $\Delta$ denotes the gap between the errors of the low-degree and high-degree bins.}
    \label{Fig: figure_qualitative_degree_appendix}
    \vskip -15pt 
\end{wrapfigure}
\subsection{Further Qualitative Analysis}
We present an additional qualitative comparison of \system{} against baseline architectures based on atom-wise E-RMSD, as shown in Figure~\ref{fig:figure_analysis}. 
Analogous to previous analyses, we calculated the atom-wise E-RMSD between the ground-truth 3D coordinates and predictions on the QM9 dataset. For an atom $i$ in the $m$-th molecule, the atom-wise E-RMSD, $\text{E-RMSD}(i)$, is formulated as:
\begin{equation}
    \text{E-RMSD}(i) = \frac{p^{(m)}}{\widehat p^{(m)}}w_i\lVert \widehat{\bm C}_i-\bm C_i\rVert_2,
\end{equation}
where $\widehat{\bm C}_i$ and $\bm C_i$ denotes the predicted and ground-truth coordinates of atom $i$. $\frac{p^{(m)}}{\widehat p^{(m)}}$ is a Boltzmann factor of the $m$-th molecule and $w_i$ denotes the normalized atom-wise force, as detailed in Section~\ref{3_analysis}.
Following the same procedure, we categorized nodes into the low-degree and high-degree groups based on the relative degree, and computed the average atom-wise RMSD for each group, as depicted in Figure~\ref{Fig: figure_qualitative_degree_appendix}. 

Consistent with the results in Section \ref{5.5}, \system{} achieves significant error reductions for both low-degree and high-degree groups.
Specifically, when compared against GTMGC, \system{} further reduces the atom-wise E-RMSD for lower-degree groups by 0.18, compared to a reduction of 0.14 for high-degree groups.
Furthermore, our method greatly reduces the gap between the atom-wise RMSD of low-degree and high-degree groups, achieving a 7.37\% percentage reduction in the gap compared to GTMGC, demonstrating the efficacy of \system{} in accurately modeling interatomic interactions.

\section{Experimental Settings}\label{C}
\subsection{Dataset Description}

We evaluate our method on three benchmark datasets: QM9, Molecule3D, and GEOM-Drugs. Details for each datasets are specified below.
\begin{itemize}
\item \textbf{QM9} is a widely-used quantum chemistry dataset containing molecular geometries, electronic properties, and energy attributes for small organic molecules with up to 9 heavy atoms. The 3D conformations are obtained using density functional theory (DFT). We adopt the data split proposed by~\citep{qm9split}.
\item \textbf{Molecule3D} is a large-scale dataset consisting of approximately 4 million molecules. Each molecule is annotated with 2D molecular graphs, ground-state 3D conformations, and various quantum properties. We follow the splits used in~\citep{gtmgc}, including a random split and a scaffold split. The scaffold split groups molecules based on their core substructures, allowing for a more realistic evaluation.
\item \textbf{GEOM-Drugs} is a subset of the GEOM dataset, focusing on drug-like molecules. For each molecule, multiple conformation sets along with their chemical properties are provided. Thus, we choose the most stable conformation with respect to Boltzmann energy for each molecule for our experiments. For data splits, we adhere to the splits used in ~\citep{ganea2021geomol}.
\end{itemize}

\subsection{Metrics Description}~\label{C.2}
We consider the following metrics for evaluation.
\begin{itemize}
    \item \textbf{MAE}. Mean absolute Error (MAE) quantify the accuracy of predicted interatomic distances relative to the ground truth on a pairwise basis. Let $\widehat{\bm{D}}_{ij}$ and $\bm{D}_{ij}$ represent the predicted and ground truth distances between atoms $i$ and $j$, respectively. The metric is formulated as follows:
    \[
\text{MAE}(\widehat{\bm{D}}, \bm{D}) = \frac{1}{N^2}\sum_{i,j\in\mathcal V} |\widehat{\bm{D}}_{ij} - \bm{D}_{ij}|\]
    \item \textbf{RMSE}. Root Mean Square Error (RMSE), also quantifies the error in interatomic distances, on a pairwise basis. Specifically, the metric is formulated as follows:
\[
    \text{RMSE}(\widehat{\bm{D}}, \bm{D}) = \sqrt{\frac{1}{N^2}\sum_{i,j\in\mathcal V} (\widehat{\bm{D}}_{ij} - \bm{D}_{ij})^2}
\]
\item \textbf{RMSD}. Root Mean Square Deviation measures the spatial deviation between two conformations that are aligned using the Kabsch algorithm~\citep{kabsch}. Only heavy atoms (\ie, excluding hydrogens) are considered during this calculation. Let $\widehat{\bm{G}}_i$ and $\bm{G}_i$ represent the aligned conformations of atom $i$. RMSD is defined as:
\[
\text{RMSD}(\widehat{\bm{G}}, \bm{G}) = \sqrt{\frac{1}{N}\sum_{i\in\mathcal V} \lVert\widehat{\bm{G}}_{i} - \bm{G}_{i}\rVert_2}
\]  

\item \textbf{E-RMSD}. Energy-weighted RMSD, considers chemical feasibility of the predictions on top of the spatial deviations. The probability likelihood of the whole preicted conformation $\widehat{\bm{G}}$ with respect to the ground state $\bm{G}$ is accounted via the Boltzmann factor (\ie 
 $\frac{p}{\widehat{p}} = \text{exp}\left(\frac{\hat{E} - E}{kT}\right)$), while the feasibility at atom-level is considered via sum-normalized force (\ie $w_i = \frac{F_i}{\sum_{j\in\mathcal V}{F_{j}}}$) acting upon each atom. Both the Boltzmann factor and force is calculated using Merck Molecular Force Field~\citep{mmff}. For all our experiments, we set $T = 298.15~\text{K}$ and $k=0.001987~\text{kcal mol}^{-1} \text{K}^{-1}$, reflecting standard laboratory conditions.

\end{itemize}

\subsection{Implementation Details}
We adopted the evaluation protocols and train/validation/test splits from \cite{gtmgc} for the QM9 and Molecule3D datasets and from \cite{jing2022torsional} for the GEOM-DRUGS dataset. Test results were derived from the best-performing model on the validation set based on the D-MAE metric. All GNN architectures were implemented using PyTorch~\cite{NEURIPS2019_pytorch} and PyTorch Geometric~\cite{2019torch_geometric}. The experiments were conducted on RTX Titan and RTX 3090 (24GB) GPU machines. Throughout all experiments except for \cite{jing2022torsional, gtmgc}, we set the hidden dimension to 512 and the number of layers to 8; for these references, we adopted their optimal  configurations. We employed the AdamW optimizer with a batch size of 100 and no weight decay. Learning rates were initially warmed up from 0 and then fixed based on the best validation performance on the QM9 dataset, within the range of [3e-5, 5e-5, 7e-5, 9e-5]. The number of attention heads was set to 8. We used a seed of 42 for all experiments and trained all models for 20 epochs, following the configuration in \cite{gtmgc}.

\end{document}